\documentclass[
    aps,
    pra,
    twocolumn,
    notitlepage,
    nofootinbib,
    superscriptaddress
]{revtex4-2}

\usepackage[utf8]{inputenc}
\usepackage[english]{babel}

\usepackage{graphicx}
\usepackage{amsmath}
\usepackage{amssymb}
\usepackage{amsfonts}
\usepackage{mathtools}
\usepackage{physics}
\usepackage{bm}
\usepackage{subcaption}
\usepackage{comment}
\usepackage{nth}
\usepackage{latexsym}
\usepackage{soul}
\usepackage{mathrsfs}
\usepackage{color}
\usepackage{hyperref}

\newcommand{\be}{\begin{equation}}
\newcommand{\ee}{\end{equation}}

\newcommand{\bear}{\be\begin{array}}
\newcommand{\bea}{\begin{eqnarray}}
\newcommand{\eea}{\end{eqnarray}}

\newcommand{\bp}{{\bf p}}
\newcommand{\bP}{{\bf P}}
\newcommand{\br}{{\bf r}}

\newcommand{\bk}{{\bf k}}

\newcommand{\expv}[1]{\left\langle #1 \right\rangle}

\newcommand{\la}{\langle}
\newcommand{\ra}{\rangle}

\newcommand{\dst}{\displaystyle}
\newcommand{\fr}[2]{\frac{{\dst #1}}{{\dst #2}}}

\begin{document}

\title{Interaction of twisted light with free twisted atoms}
\author{I.~Pavlov}
\email{ilia.pavlov23@yandex.ru}
\affiliation{School of Physics and Engineering,
ITMO University, 197101 St. Petersburg, Russia}%
\affiliation{Petersburg Nuclear Physics Institute of NRC “Kurchatov Institute”, Gatchina 188300, Russia}
\author{A.~Chaikovskaia}
\affiliation{School of Physics and Engineering,
ITMO University, 197101 St. Petersburg, Russia}%
\author{D.~Karlovets}
\email{d.karlovets@gmail.com}
\affiliation{School of Physics and Engineering,
ITMO University, 197101 St. Petersburg, Russia}%
\affiliation{Petersburg Nuclear Physics Institute of NRC “Kurchatov Institute”, Gatchina 188300, Russia}

\date{\today}
\begin{abstract}
We investigate absorption and scattering of structured light by atoms, treating the photon and the atomic center of mass as spatially localized wave packets. We show that vortex photons can transfer orbital angular momentum (OAM) to the atomic center of mass with  near-perfect efficiency in head-on collisions when the impact parameter $b$ is smaller than the atomic transverse coherence length $\sigma$, which ranges from nanometers to sub-micrometer scales. Larger offsets result in a shifted mean OAM and a finite variance, both controlled by the ratio $b/\sigma$. The wave-packet nature of light enables electronic transitions that violate standard selection rules, albeit with a clear hierarchy where the dipole transition dominates. For femtosecond pulses, the finite \textcolor{black}{spatial extent} of the photon leads to measurable shaping of the resonant absorption lines. We demonstrate a transverse recoil of the atom in a vicinity of the photonic vortex, dubbed ``the superkick'', and its dual effect - ``the selfkick'' - when an initially twisted atomic packet experiences recoil upon absorbing a gaussian photon. These phenomena are within reach of experimental capabilities using structured light in combination with cold atomic beams and ions in Penning traps, providing a route to the controlled generation and manipulation of non-gaussian atomic packets.
\end{abstract}

\maketitle

\section{Introduction}
Structured waves carrying orbital angular momentum (OAM), dubbed twisted waves, are widely studied across optics, acoustics, hydrodynamics, particle physics and atomic physics \cite{BLIOKH20171, ivanov2022promises, knyazev2018beams, forbes2021structured, roadmap_strlight, roadmap_strwaves}. They enable effects inaccessible in conventional setups such as modified selection rules in atomic transitions, altered angular distributions in scattering, novel observables that cannot be probed with the plane-wave beams, and so forth \cite{schmiegelow2016transfer, babiker2019atoms, ivanov2022promises}.

Twisted beams of individual atoms have been experimentally realized using diffraction gratings \cite{luski2021vortex}. This has stimulated theoretical studies of interactions of atoms and ions with radiation, including twisted photons \cite{ivanov2022double, maslennikov2024theoretical, baturin2024conversion}, where OAM transfer and modified angular distributions have been demonstrated. The concept of twisted atomic beams, however, dates back further: earlier  works addressed their quantum-optical properties \cite{helseth2004atomic, hayrapetyan2013bessel} and proposed generation schemes based on diffraction via optical masks \cite{lembessis2014atom, lembessis2017atomic}. Despite this developments, existing theoretical approaches still rely on treating the atom and the photon in simplified forms, and calculations where they both are described as wave packets—especially non-gaussian ones—remain absent.

Here, we develop a theory of interactions between quantized radiation and localized atomic packets, including twisted states as a special case, using second-order perturbation theory and nonrelativistic velocities beyond the dipole  and the infinitely heavy nucleus approximations. We derive absorption and scattering amplitudes and introduce a generalized cross section \cite{karlovets2020effects, JHEP2017}, revealing effects absent in the plane-wave treatments, such as coherent atomic recoil -- the so-called superkick \cite{barnett2013superweak} -- and its dual, which we term the ``selfkick'', as well as the transfer of OAM from photons to the atomic center of mass, atomic packet shaping, and absorption line distortions. These phenomena are experimentally accessible with current technologies employing femtosecond pulses and supersonic atomic beams, trap-release setups \cite{luski2021vortex, fuhrmanek2010imaging, fuhrmanek2010measurement} and trapped ions.

The natural system of units $\hbar=c=\varepsilon_0=1$ is used. The mass $m$ stands for the electron mass, while $M$ is that of the nucleus; $e<0$ is the electron charge.

\section{Interaction of an atom with external field}
We start with a theoretical framework for the interaction of localized atoms with quantum light. Specifically, we use the quantum-mechanical perturbation theory for a nonrelativistic hydrogen-like ion interacting with a quantized electromagnetic field in free space. Next, we stay within first two orders of perturbation theory enabling the description of single-photon absorption (photoexcitation) and scattering. The key differences of our approach from the standard methods are the following:
\begin{enumerate}
    \item We treat both the photon and the atom as spatially localized and nonstationary packets. The ``vortex'' packets carrying OAM are of special interest to us, and we use the basis of the Hermite-Gaussian$\times$ Laguerre-Gaussian (HG$\times$ LG) modes to describe them (see details in Supplementary Materials, Sec. I, II, III).
    \item We use neither a multipole expansion nor the infinitely heavy nucleus approximation.
\end{enumerate}

Note that the Schr\"odinger equation for an electron and a nucleus is usually separated into two equations: one with Coulomb potential for the \textit{relative motion}, which coincides with the electron in the limit of infinitely heavy nucleus, and one for the free center-of-mass (CM) dynamics (see Supplementary Materials, Sec. II). In what follows, we refer to the relative motion component as ``electron'', and sometimes refer to the CM component simply as ``atom''.

\subsection{Absorption matrix element}
Let a free hydrogen-like atom with a CM momentum-space wave function $\psi^{\text{(CM)}}_i(\bP_i)$ and the electronic state $\psi^e_i(\br)$ absorb a photon in a non-plane-wave state $|\gamma_i\ra$. \textcolor{black}{The vector $\bm r = \bm r_e - \bm r_n$ is the relative coordinate of electron w.r.t. the nucleus.} Using the perturbation theory, detailed in Supplementary Materials, Sec. V, we obtain the matrix element of transition to a final electronic state $\psi^e_f(\br)$ and a CM state $\psi_{f}^{\text{(CM)}}({\bP}_f)$:

\begin{multline}
\label{S_with_eJ}
S_{fi}^{(1)} = 
-ie\sum\limits_{\lambda_i=\pm 1}\int \fr{d^3P_i}{(2\pi)^3}\,\fr{d^3P_f}{(2\pi)^3}\,\fr{d^3k_i}{\sqrt{2\omega_{\bk_i}}(2\pi)^3}\\
\times (2\pi)^4\delta \left(\varepsilon^{\text{(CM)}}_i(\bP_i)+\varepsilon^{e}_i+\omega_i-\varepsilon^{\text{(CM)}}_f(\bP_f)-\varepsilon^e_f\right)\\
\times \delta(\bP_i+\bk_i- \bP_f)\,\left(\psi_{f}^{\text{(CM)}}({\bP}_f)\right)^*\, {\bm e}_{\lambda_i}(\bk_i)\cdot{\bm J}_{if}({\bk}_i, \bP_i) \\
\times \psi^{\text{(CM)}}_i({\bP}_i)\psi^{\gamma}_i({\bk}_i,\lambda_i)e^{-i\bk_{i\perp} \bm{b}}.
\end{multline}

Here, ${\bm e}_{\lambda_i}(\bk_i)$ is a polarization vector of a plane-wave mode (${\bm e}_{\lambda_i}(\bk_i)\cdot \bk_i = 0$ in the Coulomb gauge), $\lambda_i=\pm1$ is the plane-wave mode helicity, $\psi^{\gamma}_i(\bk_i,\lambda_i) \equiv \la\bk_i,\lambda_i|\gamma_i\ra$ is the photonic wave function, and $\bf b$ is the impact parameter between the photon propagation axis and the CM. $\varepsilon^e_{i, f}$ are the electronic energies, $\omega_i=|\bk_i|$ is photonic dispersion and $\varepsilon_i^{\text{(CM)}}(\bP_i) = \bP_i^2/(2(m+M))$ is the atomic one. The electronic transition current is denoted as
\begin{multline}
\label{Current}
{\bm J}_{if}({\bk}_i, \bP_i) = -\fr{1}{m}\,{\bm j}^{(1)}_{if}-\fr{1}{m+M}\,{\bm j}^{(2)}_{if}\\
-\fr{Z}{M}\,{\bm j}^{(3)}_{if}+\fr{Z}{m+M}\,{\bm j}^{(4)}_{if},
\end{multline}
and
\bea
\label{currents}
{\bm j}^{(1,2,3,4)}_{if} = \int d^3r\,(\psi_{f}^e(\br))^* 
\begin{Bmatrix}
e^{i\bk_i\cdot\br\frac{M}{m+M}}\hat{\bp}_r \\[2pt]
e^{i\bk_i\cdot\br\frac{M}{m+M}}\bP_i \\[2pt]
e^{-i\bk_i\cdot\br\frac{m}{m+M}}\hat{\bp}_r \\[2pt]
e^{-i\bk_i\cdot\br\frac{m}{m+M}}\bP_i
\end{Bmatrix}
\psi^e_i(\br)
\eea
where $\hat{\bp}_r = -i \fr{\partial}{\partial r}$. The structure of this current is defined by the general interaction Hamiltonian derived in Sec. IV  of Supplementary Materials. The first term ${\bm j}^{(1)}_{if}$ corresponds to electronic transitions caused by the interaction of what we call ``electronic'' component with photons; ${\bm j}^{(4)}_{if}$ represents the effect of the nucleus on the CM interaction; ${\bm j}^{(2)}_{if}$ and ${\bm j}^{(3)}_{if}$  are the ``cross-terms'' responsible for the coupling of the photon with the electronic part of CM and with the nuclear part of the relative motion, respectively. The first current dominates in most cases, exceeding the others by a factor of at least $M/m$, which is roughly 2000 for hydrogen and can reach $10^5$ for heavy elements with $Z \gg 1$. Moreover, the second and the fourth currents {\it completely vanish in the dipole approximation}, when $\bk_i\cdot \br \to 0$, because the electronic state must be excited in the absorption process and the initial and final states are orthogonal.

The calculations can be simplified in a coordinate system with $z$ axis aligned with \textcolor{black}{$\bk_i$, which means that different frames are used for every plane wave component of the photon wave packet.} The electronic wave functions are transformed under this rotation with the aid of Wigner d-matrices $d^\ell_{m m'}$ (see Supplementary Materials, Sec. VI), and the dot product in Eq. \eqref{S_with_eJ} can be rewritten as ${\bm e}_{\lambda_i}(\bk_i)\cdot{\bm J}_{if}({\bk}_i, \bP_i) \equiv  e^{i(m_i-m_f)\phi_k} \mathcal J_{\lambda_i}(\bk_i, \bP_f)$ with 

\begin{multline}
     \mathcal J_{\lambda_i}(\bk_i, \bP_f) = \sum_{m_i', m_f'}  d^{\ell_i}_{m_im_i'}(\theta_k) d^{\ell_f}_{m_fm_f'}(\theta_k)\Big(-\fr{1}{m}\,{j}^{(1)}_{\lambda}\\
     -\fr{1}{m+M}\,{j}^{(2)}_{\lambda}-\fr{Z}{M}\,{j}^{(3)}_{\lambda}+\fr{Z}{m+M}\,{j}^{(4)}_{\lambda}\Big),
\end{multline}
where $m_i$ and $m_f$ are initial and final magnetic quantum numbers of the electron and $j^{(1,2,3,4)}_{\lambda_i}$ are the spherical components of the currents defined in Eq.~(S101).

\subsection{The state of the center-of-mass}
\label{Sec:state of CM}

Specifying a final state of the center-of-mass $\psi_f^{\text{(CM)}}(\bP_f)$ after absorption implies that this state is \textit{projectively measured}. One can, however, readily derive \textit{an evolved state} \cite{EPJC2023, EPJC2024} of the atomic CM arising as a result of the absorption of the photon, which does not depend on the detection scheme, as $|\psi^{\text{(CM, ev)}}(t_f)\ra = \la \psi^e_f(t_f);\gamma_f(t_f)|\hat{S}|\Psi_i(t_i)\ra$.
The wave function of this evolved state $\psi^{\text{(CM, ev)}}(\bP_f) \equiv \la\bP_f|\psi^{\text{(CM, ev)}}(t_f)\ra$  is essentially the \textit{amplitude} of CM transition from the initial state to a plane wave:
\begin{multline}
    \label{Ev_state_CM}
\psi^{\text{(CM, ev)}}(\bP_f) =  - ie\sum\limits_{\lambda_i}\int \fr{d^3P_i}{(2\pi)^3}\fr{d^3k_i}{\sqrt{2\omega_{\bk_i}}(2\pi)^3} \\
\times (2\pi)^4\delta \left(\varepsilon^{\text{(CM)}}_i(\bP_i)+\varepsilon^{e}_i+\omega_i-\varepsilon^{\text{(CM)}}_f(\bP_f)-\varepsilon^e_f\right)\cr
\times \delta(\bP_i+\bk_i- \bP_f)\, e^{i(m_i-m_f)\phi_k-i\bk_\perp \bm b}  \mathcal J_{\lambda_i}(\bk_i, \bP_f) \\
\times \psi^{\text{(CM)}}_i({\bP}_i)\psi^{\gamma}_i({\bk}_i,\lambda_i)\,e^{-i t_f \varepsilon^{\text{(CM)}}_f(\bP_f)}.
\end{multline}
The phase dependence on $t_f$ here indicates the free evolution of the CM wave function, and thus the probability density of the evolved state in the momentum space does not depend on time.

Focusing now on the OAM transfer, let us now assume that both the incoming photon and the initial CM state are twisted and single out the phase factors indicative of the angular momentum:
\begin{gather}
    \psi_i^{\text{(CM)}}(\bP_i) = \tilde\psi_i^{\text{(CM)}}(P_{i \perp}, P_{iz})e^{i\ell^{(\text{CM})}_i \phi_i}\\
     \psi_i^{\gamma}(\bk, \lambda_i) =  \tilde\psi^{\gamma}(k_{\perp}, k_{z}, \lambda_i)e^{i(\ell_\gamma+\lambda_i) \phi_k},
\end{gather}
where $\ell^{(\text{CM})}_i = 0, \pm 1, \pm 2,..., \ell_\gamma = 0, \pm 1, \pm 2,...$ are the \textit{orbital} angular momenta of the CM and of the photon, respectively, $\lambda_i$ is the photon helicity. By rewriting the transverse delta function in cylindrical coordinates (see Sec. VII of the Supplementary Materials for details), one can explicitly verify that the evolved state is an eigenfunction of the OAM z-projection operator:
\begin{eqnarray}
\label{eigenfunc}
    -i \fr{\partial}{\partial \phi_f} \psi^{\text{(CM, ev)}}(\bP_f) = \ell_0 \:\psi^{\text{(CM, ev)}}(\bP_f),
\end{eqnarray}
where $\ell_0 =\ell_\gamma+\lambda_i+\ell^{(\text{CM})}_i+m_i-m_f$.


As expected, cylindrical symmetry gives rise to the conservation of the TAM $z$-projection. In realistic experimental conditions, it is very challenging to control the impact-parameter $\bm b$. Consequently, in reality the evolved state of the CM can become {\it a superposition of twisted states} with the different OAM values. This follows from expanding the $\bm b$-dependent phase in Eq. \eqref{Ev_state_CM} using the Jacobi–Anger formula $e^{-i\bm k_\perp \bm b} = \sum \limits_{m=-\infty}^{+\infty} (-i)^m J_m(k_\perp b)e^{i m (\phi_k-\phi_b)}$,
where $J_m$ is the Bessel function, so the evolved CM state becomes a superposition of states with OAM $\ell_0 \pm m$. Since the realistic transverse coherence length can be estimated as $\sigma_\perp^\gamma \sim k_\perp^{-1} \approx \;1–10\; \mu \text{m}$, the quantity $k_\perp b$ can be considered a small parameter as $b \ll 1$–$10$ $\mu$m aligns with typical beam positioning accuracy in trapped-atom experiments \cite{schmiegelow2016transfer, schmiegelow2022, afanasev2018experimental}. \textcolor{black}{Note that, throughout this work, the term ``coherence length'' is used synonymously with the spatial extent of the packet, for both photonic and atomic packets.} Thus, the asymptotic form of Bessel function can be used: $J_m(\kappa b) \approx (\kappa b/2)^m/m!$. Although the Jacobi-Anger expansion now suggests that OAM sidebands $\ell_0 \pm m$ are suppressed by $(k_\perp b)^m\ll 1$, the actual OAM distribution is {\it not} symmetric. This is due to the evolved state being not normalized to unity: its norm reflects the absorption probability that depends on $\ell_\gamma$, favoring {\it smaller} OAM values.

This $\ell_\gamma$-dependence can be understood by considering a gaussian CM wave function. In this case, the integrand in Eq. \eqref{Ev_state_CM} consists of a slowly varying function of $\phi_k$— the main dependence of which on $\phi_k$ comes precisely from the CM wave function — multiplied by $e^{i(\ell_0+m)\phi_k}$. Integrating over $d^3 P_i$ and applying the delta function yields $P_{i\perp} = \sqrt{P_{f\perp}^2 + k_{\perp}^2 - 2P_{f\perp}k_{\perp} \cos(\phi_k-\phi_p)}$. The $\phi_k$ integral then reduces to
\begin{multline}
\int_0^{2\pi} e^{(\sigma^{(\text{CM})}_\perp)^2 P_{f\perp}k_{\perp} \cos(\phi_k-\phi_p)} e^{i(\ell_0+m) \phi_k} \frac{d \phi_k}{2\pi}\\ \propto I_{\ell_0+m} \left((\sigma^{(\text{CM})}_\perp)^2 P_{f\perp}k_{\perp}\right),
\end{multline}
where $I_m$ is a modified Bessel function.
Since characteristic transverse momenta are defined by inverse coherence lengths, the evolved CM state can be estimated as
\begin{multline}
\label{estimated_dist}
\ket{\psi^{\text{(CM, ev)}}} \propto 
\sum \limits_{m=-\infty}^{\infty}(-i)^m J_m(b / \sigma_\perp^\gamma) \\
\times I_{\ell_0+m} \left(\sigma^{(\text{CM})}_\perp / \sigma_\perp^\gamma \right) \ket{\ell_0+m}.
\end{multline}

\begin{figure}[h]
    \centering
    \includegraphics[width=0.8\linewidth]{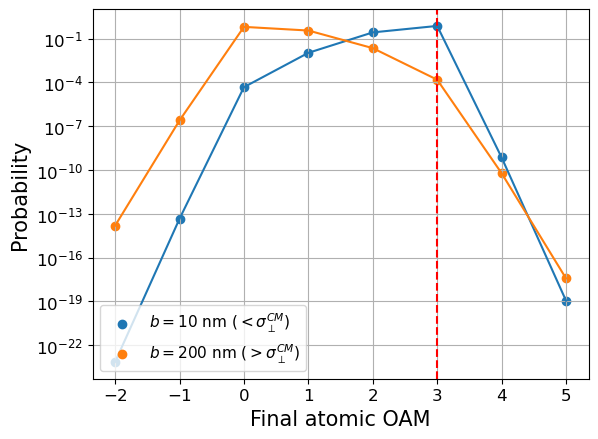}

    \caption{Estimated probability distribution \eqref{estimated_dist} of OAM in the evolved state of CM. Initial atomic packet width is $\sigma_\perp^{(\text{CM})} = 50$ nm,  photonic packet width $\sigma_\perp^{\gamma} = 1$ $\mu$m, transferred OAM at zero $b$ equals $\ell_0 = 3$ (indicated by red line). For $b = 10$ nm: $\langle \ell^{\text{(CM)}}\rangle = 2.72$, $\Delta \ell^{\text{(CM)}} = 0.47$. For $b = 200$ nm: $\langle \ell^{\text{(CM)}}\rangle = 0.40$, $\Delta \ell^{\text{(CM)}} = 0.53$.}
    \label{OAM_distrib}
\end{figure}

Two examples of this distribution are shown in Fig. \ref{OAM_distrib}. We explain them analytically in Sec.~VIII of Supplementary Materials by using the asymptotics of Bessel functions. Interestingly, the angular dependence of the probability density for the two evolved states in Fig.~\ref{OAM_distrib} is nearly identical despite their very different OAM spectra. Indeed, the superposition of two dominating neighboring modes, which is proportional to  $c_1(-i)^{-3} + c_2(-i)^{-2} e^{i\phi}$ in one case and to $c_1 e^{3i\phi} + c_2(-i)^{-1} e^{2i\phi}$ in the other, leads in both cases to the same azimuthal dependence of the probability density $\propto 1 + 2 c_1 c_2 \sin\phi$.

\subsection{Second-order processes: scattering}
\label{second_order}

Analogously to absorption, we generalize the well-known resonant scattering amplitude \cite{messiah2014quantum, sakurai1967advanced}, arising from the second-order perturbation theory, to our problem. The details of derivation are given in Supplementary Materials, Sec. V, and the plane-wave amplitude is given by Eq.~(S88).


There is, however, another contribution to the scattering, which arises from the Hamiltonian quadratic in the vector potential in the first order of perturbation theory (the ``seagull'' diagram \cite{sakurai1967advanced}). The corresponding plane-wave transition amplitude~(S90) is denoted as $S_{fi}^{(1.5, PW)}$. In principle, the first-order and second-order scattering amplitudes interfere with each other. If there is no resonance in the system (e. g. for scattering by a free electron), the contribution of $ S_{fi}^{(2)}$ to the total amplitude is known to be suppressed w.r.t. $ S_{fi}^{(1.5)}$ by a factor of $\omega / m$ \cite{messiah2014quantum}. In the resonant regime, however, the situation becomes {\it completely reversed}, and we can safely neglect the ``seagull'' term $ S_{fi}^{(1.5)}$. A straightforward generalization for wave packets in the initial state reads
\begin{equation}
\label{S_1.5_2_packets}
    S_{fi} = \sum_{\lambda_i} \int \fr{\dd^3 P_i}{(2\pi)^3} \fr{\dd^3 k_i}{(2\pi)^3}  S_{fi}^{(PW)} \psi^{\text{(CM)}}_i({\bP}_i)\psi^{\gamma}_i({\bk}_i,\lambda_i),
\end{equation}
where the plane-wave scattering amplitude can be either ~(S88), or~(S90), or their sum.
By analogy with the absorption amplitude \eqref{Ev_state_CM}, the S-matrix element \eqref{S_1.5_2_packets} can be regarded as a two-particle momentum-space wave function describing an \textit{entangled evolved state} of the photon and the atom after the scattering process ending at the time instance $t_f$:
\begin{equation}
    S_{fi} e^{-i t_f (\varepsilon^{\text{(CM)}}_f(\bP_f) + \omega_f)} = \la\bP_f, \bk_f, |\psi^{\text{(CM+$\gamma$, ev)}}(t_f)\ra
\end{equation}
For a head-on collision, this evolved state must be an eigenfunction of the \textit{total} angular momentum (TAM) projection operator that is a sum of the TAM operators for each counterpart, the CM and the photon. The individual values of the photon TAM and the electron OAM, however, are not defined due to entanglement, which is why the possibility of the vortex photon/atom generation depends on the measurement schemes \cite{EPJC2023, EPJC2024, chaikovskaia2024vavilov}.

\subsection{Probabilities and cross sections}
In contrast to the plane-wave scattering theory \cite{sakurai1967advanced, messiah2014quantum}, current more general wave-packet approach makes \textit{the total probability} of absorption or scattering a well-defined quantity. Indeed, the total probability of absorption can be evaluated by integrating the evolved state probability density over all possible final momenta of the CM $\bP_f$,
\begin{equation}
\label{prob_abs}
    P_{\text{abs}} = \int \fr{\dd^3 P_f}{(2\pi)^3} |\la\bP_f|\psi^{\text{(CM, ev)}}\ra|^2 
\end{equation}
For the scattering probability, an additional integration over the final momenta of the photon and summation over its polarizations is needed,
\begin{equation}
\label{prob_sc}
   P_{\text{scatt}} = \sum_{\lambda_f}\int \fr{\dd^3 P_f}{(2\pi)^3} \fr{\dd^3 k_f}{(2\pi)^3} \Big| S_{fi}^{(2)} +  S_{fi}^{(1.5)}\Big|^2 .
\end{equation}
We compute both probabilities \eqref{prob_abs} and \eqref{prob_sc} numerically, employing the Monte-Carlo method to evaluate the six-dimensional integral in Eq.\ \eqref{prob_sc}.

Although the probability for collision of wave packets is a well defined quantity, it cannot be directly compared to a standard plane-wave model since the most common observable is a cross section. A general expression for a cross-section for an arbitrary scattering process with wave packets was derived in Ref. \cite{JHEP2017, karlovets2020effects} extending  the earlier work \cite{kotkin1992processes}. It was used for some practical calculations, for instance, in \cite{liu2023threshold, li2024unambiguous, liu2026vortex, shchepkin2025absorption}. Here we also follow this approach, defining the generalized cross-section as
\begin{equation}
\label{gen_sigma}
    \sigma_{\text{abs}, \text{scatt}} = \frac{P_{\text{abs}, \text{scatt}}}{L},
\end{equation}
where $L$ is \textit{luminosity} characterizing the spacetime overlap between the photonic and atomic (CM) packets, which in the paraxial approximation looks as
\begin{equation}
\label{lum}
    L = |v_1-v_2| \int \dd^3r \dd t \:|\psi^{\text{(CM)}}_i(\br, t)|^2|\psi^{\gamma}_i(\br, t)|^2,
\end{equation}
where $\psi^{\text{(CM)}}_i(\br, t)$ and $\psi^{\gamma}_i(\br, t)$ are the spatial wave functions of the atom and of the photon, respectively, where the latter is related to the energy density of the photon \cite{Mandel_Wolf_1995}. The factor $|v_1-v_2|$ is a relative velocity, which in the paraxial approximation can be computed via the average momenta of the two packets. Beyond the paraxial case, a more general form of the luminosity should be used \cite{JHEP2017, karlovets2020effects}. The spatial CM wave function is defined as
\begin{equation}
\label{position_psi}
    \psi^{\text{(CM)}}_i(\br, t) \equiv \int \fr{\dd^3 P}{(2\pi)^3} \psi^{\text{(CM)}}_i(\bP) \:e^{i \bP \cdot \br -i\varepsilon^{\text{(CM)}}_i(\bP)t}.
\end{equation}
Although the very definition of the spatial wave function - especially a scalar one - is a bit tricky for photon \cite{Mandel_Wolf_1995, Scully}, for a paraxial and quasi-monochromatic packet with a fixed helicity $\lambda$ it can be defined analogously to Eq. \eqref{position_psi}  (see Sec. III of Supplementary Materials for details) as

\begin{figure}
	\centering
		\includegraphics[width=\linewidth]{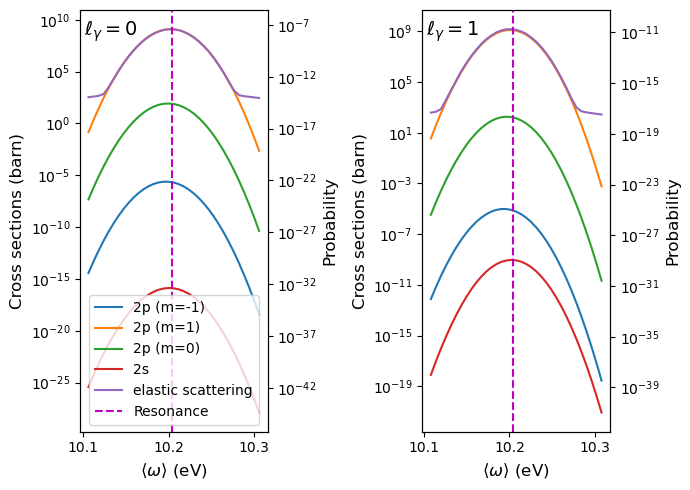}

	\caption{$n=1\to n=2$ transitions in hydrogen: the probabilities \eqref{prob_abs} and cross sections \eqref{gen_sigma} of absorption  versus the \textit{average} photon energy.
    $\expv{\omega} \equiv \sqrt{\expv{k_z^2} + (2n_\gamma + |\ell_\gamma|+1)/(\sigma_\perp^\gamma)^2}$ for different values of the photon OAM $\ell_\gamma$. The photon packet~(S35) has $\lambda_i=1$, $\sigma^\gamma_z=10$ $\mu$m, $\sigma^\gamma_\perp=1$ $\mu$m, the radial and longitudinal indices are $n_\gamma=k_\gamma=0$. The CM wave function~(S9) is gaussian with $\sigma^{\text{(CM)}}_\perp = \sigma^{\text{(CM)}}_z = 20$ nm, and it has a vanishing mean momentum, $\expv{P_{iz}} = 0$. Note that the transverse rms size of the photonic packet equals $\sigma_\perp^\gamma\sqrt{2n_\gamma+|\ell_\gamma|+1}$ rather than simply $\sigma_\perp^\gamma$.}
    \label{n1_to_n2}
\end{figure}

\section{Results}
\subsection{General features of cross sections} \label{general}
In Fig.~\ref{n1_to_n2}, we show numerical results for the absorption and scattering cross sections for a hydrogen atom when the photon frequency is close to $1s \to 2p$ resonance (around $10.2\,\text{eV}$). These transitions connect the ground state $1s$ to the first excited states with the principal quantum number $n=2$, which include the $2s$ and $2p$ levels. We neglect the fine-structure splitting within the $n=2$ manifold, treating these levels as degenerate for simplicity. Throughout this section, we restrict ourselves to collisions with a vanishing impact parameter $b\to 0$. Fig.~\ref{n1_to_n2} also displays the total probabilities on the second vertical axis. This representation is valid because the luminosity depends only weakly on the average photon energy, $\expv{\omega}$. 

First, although the natural linewidth is not taken into account in the absorption amplitude, the shape of the delta-like ``absorption line'' is regularized due to the  photon's lack of monochromaticity. For the chosen parameters, such {\it instrumental broadening is much more significant} than the radiative one: a femtosecond photon has a typical longitudinal coherence length of $\sigma_z^\gamma \sim 10 \; \mu m$ corresponding to the spectral bandwidth of the order of $0.02$ eV, whereas the natural linewidth of the hydrogen $2p$ state is approximately $4\cdot10^{-7}$ eV. 

Second, we observe the {\it violation of the conventional OAM selection rules} \cite{schmiegelow2016transfer, afanasev2018experimental, alharbi2023photoexcitation, zanon2023engineering}: transitions from the $1s$ state to \textit{all} the states with $n=2$ are in principle possible, whereas in the plane-wave regime for helicity $\lambda$ only the transition to $2p$ state with $m_f=\lambda$ would be allowed. Due to the OAM conservation, accounting for the CM is essential for all amplitudes to be nonvanishing. Transitions to $m_f\neq \lambda$ levels, however, become possible primarily because of the momentum spread in the absorbed photon packet: each plane-wave component of the photon effectively couples to the ``rotated'' electronic wave function~(S94), which contains a contribution from the $m_f=\lambda$ state. In contrast, the single-photon $1s-2s$ transition is only possible thanks to the non-dipole corrections and the interaction with the CM, as the transition currents $j_{\lambda_i}^{(1, 3)}$ vanish in this case (see Table~$S1$). However, there is an obvious hierarchy of the probabilities. For instance, for a gaussian photon, each successive transition in the hierarchy is suppressed by some 10 orders of magnitude, and the dipole transition with $m_f=\lambda = \pm 1$ remains the dominant one even for the absorbed photon carrying nonzero OAM.

Importantly, the scattering cross section coincides with the absorption cross section of the most probable transition in a small vicinity of the resonance, whereas outside this region they have different asymptotics - gaussian and polynomial decay for absorption and scattering, respectively. The qualitative explanation of this behavior is the following. Absorption occurs only to those plane waves that are strictly at the resonance, and since the photonic packet (and hence its spectrum) has gaussian tails, the probability decays accordingly. Scattering, in contrast, occurs at all frequencies: if the whole photonic packet is outside the resonance, then the scattering probability has the well-known Lorentz-like asymptotics $P_{\text{scatt}} \propto 1 / (\expv{\omega} - \omega_0)^2$, where $\omega_0$ is the resonance frequency. However, if some of the plane waves turn out to be at the resonance, the scattering amplitude increases by many orders of magnitude, though stays finite because of the wave-packet averaging.

Finally, similarity of the scattering and absorption probabilities at the resonance indicates that the scattering  can be interpreted as two sequential processes: photon absorption and spontaneous emission, where the latter occurs with the total probability of $1$ as we consider the infinite timespan. This interpretation is justified if the collision time greatly exceeds the lifetime of the excited electronic state \cite{sakurai1967advanced}.

\begin{figure}
    \centering
    \includegraphics[width=0.9\linewidth]{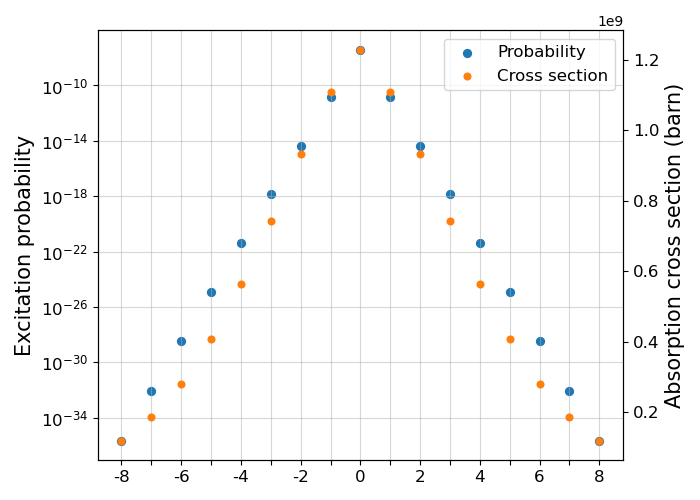}

    \caption{Probability \eqref{prob_abs} (blue points) and cross section \eqref{prob_sc}(orange points) of the dipole transition $1s\to 2p, m_f=\lambda=1$ as a function of the photon OAM $\ell_\gamma$. The probability of transitions that violate the standard selection rule decreases exponentially with $\ell_\gamma$, while the cross section decreases only linearly.}
    \label{Prob_ell_exponent}
\end{figure}

Fig.~S1 in Supplementary Materials illustrates the same as Fig. \ref{n1_to_n2}, but for $n=1 \to n=3$ transitions in hydrogen. There the situation remains qualitatively the same, apart from the increased number of available final electron states. Importantly, even for an absorbed vortex photon with $\ell_\gamma=1$ the dominant transition {\it is still the dipole one} with $m_f=\lambda = \pm 1$ and {\it not} the transition $1s \to 3d$ with $m_f=2$ although the former violates the OAM conservation (excluding CM) and the latter does not. For $\ell_\gamma=1$, the  cross section value for the most probable transition is roughly the same as that for a gaussian photon with $\ell_\gamma=0$, while the hierarchy of the suppressed transitions is slightly altered by the increase of $1s\to3d \:(m_f=2)$ and  $1s\to3s$ cross sections. For larger photon OAM values, the probability decreases exponentially, as shown in  Fig.\ref{Prob_ell_exponent}. Interestingly, the corresponding cross section decreases only linearly, since the exponential decay of probability is compensated by a comparable factor in the luminosity, caused by a vanishing field of the twisted photon around the $z$ axis.

Focusing now on the motion of the atomic CM, we conclude that twisted photons are capable of transferring their quantized OAM to the CM, \textit{thereby effectively “twisting” the atomic packet as a whole}. Importantly, this atomic OAM is {\it intrinsic and quantized}, so it does {\it not} result in any rotation of the CM around an external axis, studied in Ref.\cite{melezhik2025acceleration}. The most probable value of the transferred OAM exactly matches $\ell_\gamma$, reflecting the dominance of the dipole interaction.
Fig.~S2 in Supplementary Materials shows the same transitions, resolved by the final angular momentum of the CM. The corresponding cross section is almost the same for $\ell_\gamma=1$ and $\ell_\gamma=2$, whereas the probabilities differ by some 3 orders of magnitude. Although the cross-section is a natural observable, it is the transition probability that measures the effectiveness of OAM transfer in terms of the number of “twisted” atoms produced per unit time. For a fixed photon flux, the rate of OAM transfer will then decrease exponentially, as shown in Fig. \ref{Prob_ell_exponent}, since the inverse of the transition probability represents the number of photons required to induce a single transition.

Another factor that limits the rate of OAM transfer is the finite lifetime of the excited electronic state. The main decay channel is the electronic transition rather than the CM transition, since the coupling of the CM to the electromagnetic field is much weaker and {\it completely vanishes} in the dipole approximation for an infinitely heavy nucleus with $Z=1$. Such spontaneous emission generally produces an entangled state of the CM and the emitted photon \cite{fedorov2005spontaneous}. Whether the twisted CM state is preserved after spontaneous decay depends on the detection method \cite{EPJC2023, EPJC2024}. Measuring the emitted photon in the OAM basis keeps the atomic CM in the vortex state with a definite OAM, whereas the usual measurement in the plane-wave basis erases information about the atom’s angular momentum. 

If the photon remains undetected, the result is a mixed state of the atom and photon, which can no longer be regarded as a pure vortex state of the CM either. Therefore, experiments aiming to study \textit{pure} twisted atomic states generated by vortex photons must be performed within the lifetime of the excited electronic state. For example, in hydrogen the lifetime of the $2p$ state is of the order of a nanosecond, which imposes a significant limitation on the practical feasibility of such schemes.

All the results in this section have been presented for the hydrogen atom, but \textit{hydrogen-like ions} show essentially the same behavior. The effect of the increased nuclear mass is negligible: it enters only through terms suppressed by at least a factor of $m/M$, producing only a tiny additional decrease in the probability. The influence of the nuclear charge $Z>1$ is more subtle. At a resonance, the photon energy $\omega_{\mathbf{k}_i}$ is approximately equal to the electronic transition energy $\varepsilon^e_f - \varepsilon^e_i$, which scales as $Z^2$. Meanwhile, in the integrand of Eq.~\eqref{Ev_state_CM} the current $\mathcal J_{\lambda_i}$ scales roughly as $Z^1$. The dependence of the luminosity on $Z$ via the mean photon energy is also negligible. Consequently, the absorption cross section depends only weakly on $Z$, as shown in Fig.~S3 in the Supplementary Materials. The remaining $Z$-dependence originates from non–plane-wave effects and non-dipole corrections, since the scaling $\mathcal J_{\lambda_i} \sim Z$ holds only in the dipole approximation.

\subsection{Superkick effect}
\begin{figure*}[t]
	\centering
	\begin{subfigure}{0.325\linewidth}
		\includegraphics[width=\linewidth]{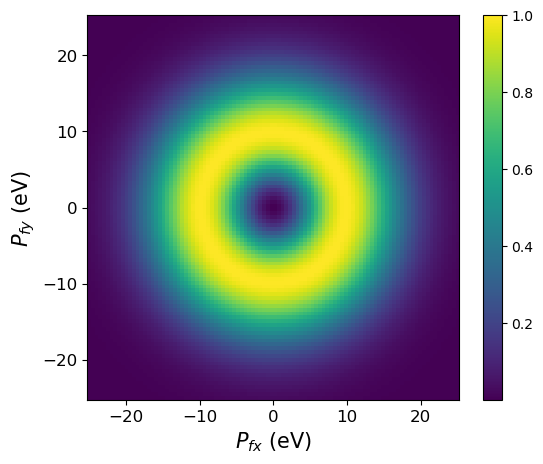}
        \caption{$b = 0$ nm;\newline
        $\expv{\ell^{\text{CM}}}=1$, $\Delta \ell^{\text{CM}}=0$.}
	\end{subfigure}
    \begin{subfigure}{0.325\linewidth}
    	\includegraphics[width=\linewidth]{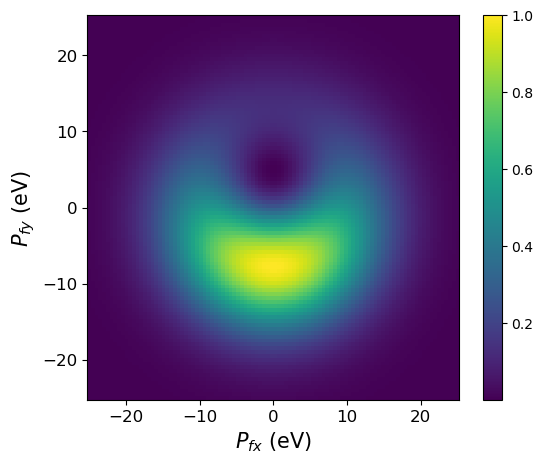}
        \caption{$b = 10$ nm; $\expv{\ell^{\text{CM}}}\approx0.8$, $\Delta \ell^{\text{CM}}\approx 0.4$.}
	\end{subfigure}
    \begin{subfigure}{0.325\linewidth}
    	\includegraphics[width=\linewidth]{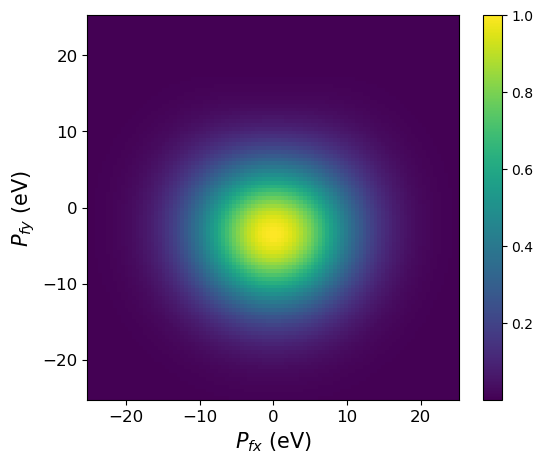}
        \caption{$b = 50$ nm; $\expv{\ell^{\text{CM}}}\approx0.14$, $\Delta \ell^{\text{CM}}\approx0.345$.}
	\end{subfigure}

	\caption{The superkick effect: the transverse probability density of the CM evolved state \eqref{Ev_state_CM} in momentum space at $P_{fz} = \expv{k_z}$. The parameters of the packets are the same as in Fig. \ref{n1_to_n2}, the dominating dipole transition is implied. The photon's OAM $\ell_\gamma=1$, the impact parameter $\bm b$ is directed along the x axis. The probability density in each figure is normalized by its maximal value.}
    \label{superkick}
\end{figure*}

\begin{figure*}[t]
	\centering
	\begin{subfigure}{0.325\linewidth}
		\includegraphics[width=\linewidth]{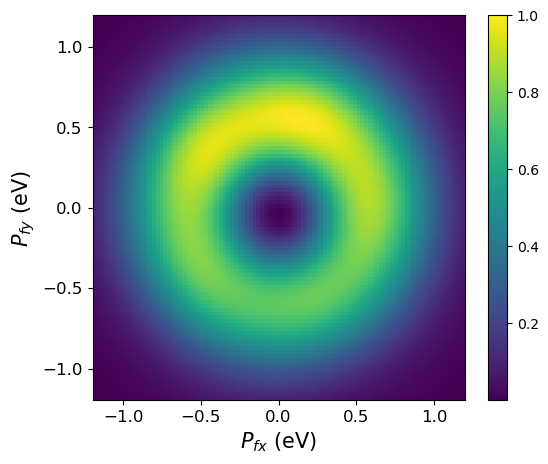}
        \caption{$b = 50$ nm}
	\end{subfigure}
    \begin{subfigure}{0.325\linewidth}
    	\includegraphics[width=\linewidth]{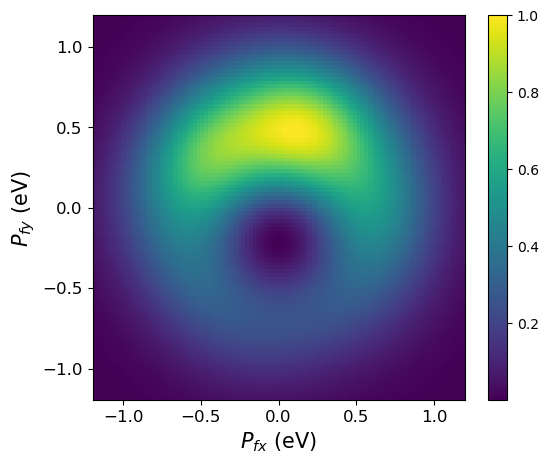}
        \caption{$b = 250$ nm}
	\end{subfigure}
    \begin{subfigure}{0.325\linewidth}
    	\includegraphics[width=\linewidth]{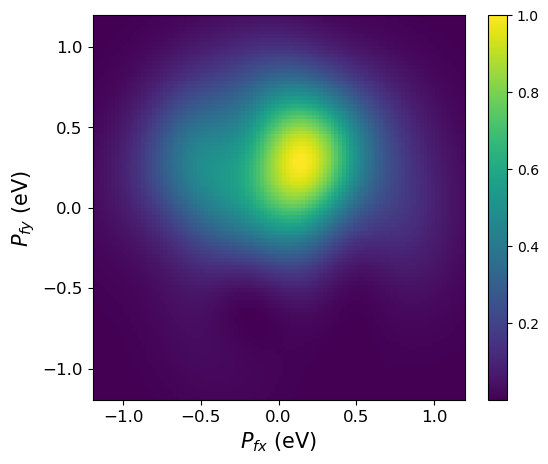}
        \caption{$b = 1000$ nm}
	\end{subfigure}

	\caption{The selfkick effect: the same as in Fig. \ref{superkick}, but for a gaussian photon and the twisted atom with the OAM $\ell_i^{\text{CM}}=1$; the atomic packet width is $\sigma_\perp^\text{(CM)}=500$ nm, the photonic transverse coherence is $\sigma_\perp^\gamma=300$ nm.}
    \label{selfkick}
\end{figure*}

Recent studies have shown that any collision between a vortex particle and a non-vortex one gives rise to an intricate kinematical effect known as the superkick \cite{barnett2013superweak, ivanov2022observability, liu2023threshold}, regardless of the specific interaction between the particles. This phenomenon manifests as an unexpectedly large transverse momentum acquired by the non-vortex packet when it is shifted transversely from the center of the incoming vortex \cite{afanasev2021recoil, afanasev2022superkicks, li2024unambiguous, liu2025superkick}. This transverse momentum is associated with the \textit{local} momentum density in the vicinity of the vortex, which can significantly exceed the mean transverse momentum of the vortex state itself. Although the momentum conservation law in this case seems to be violated, it holds for every plane-wave component comprising the colliding packets, so no paradox appears. As we address the same problem as analyzed in the original work by Barnett and Berry in Ref. \cite{barnett2013superweak} — namely, the absorption of a photon by a localized atomic packet — our model is expected to predict the same superkick effect. Importantly, our approach is fully quantum, whereas the optical vortex field was treated classically in Ref. \cite{barnett2013superweak}.

Fig. \ref{superkick} illustrates the essence of the superkick effect. For a vanishing impact parameter, the transverse structure of the photonic vortex packet is imprinted onto the CM wave function (see Sec IX of Supplementary Materials), resulting in a doughnut-shaped probability density. The characteristic transverse momentum in this case is determined by the initial momentum spread of the CM (for $\sigma_\perp^{(\text{CM})} = 20$ nm, this is $1 / \sigma_\perp^{(\text{CM})} \approx 10$ eV), rather than by the photonic transverse momentum $\sim 1 / \sigma_\perp^\gamma \approx 0.2$ eV. A transverse shift of the photon propagation axis along the $x$ direction induces a redistribution of CM momenta along the orthogonal axis: in Fig. \ref{superkick} (b), the ring-shaped distribution becomes skewed in the $y$ direction. 

Reversing the sign of the photonic OAM would induce an opposite shift due to the corresponding change in the direction of the local wave vector \cite{barnett2013superweak}. For larger impact parameters, comparable to the width of the atomic packet, the ring-shaped structure disappears entirely, Fig. \ref{superkick} (c). Such a ``kicked'' CM wave packet is clearly a superposition of states with the different OAM values rather than a single vortex state due to the broken axial symmetry. To estimate the rate of the OAM nonconservation, we numerically calculate the mean OAM of the evolved state $\expv{\ell^{\text{CM}}}$ and its standard deviation $\Delta \ell^{\text{CM}}$ (see the caption). These values are perfectly consistent with the OAM distribution derived in Sec. \ref{Sec:state of CM}.

\subsection{Selfkick effect}

Since the superkick is a purely kinematical effect, one can anticipate {\it a dual phenomenon} in the opposite configuration — namely, in a collision between an initially twisted atom and a non-vortex (gaussian) photon packet. In our simulations, we indeed observe a comparable redistribution of momenta in the CM packet, shown in Fig. \ref{selfkick}. We refer to this as {\it ``the selfkick'' effect}. The relevant parameter regime, however, differs from that of the conventional superkick. The selfkick becomes significant only for impact parameters comparable to the width of the photon packet - that is, at least hundreds of nanometers - and it becomes more pronounced as the CM packet expands. Qualitatively, this can be understood as follows: only a portion of the extended vortex packet, where the azimuthal momentum density is large, effectively interacts with the tightly focused field. As a result, the transferred momentum is redistributed in an asymmetric manner across the packet. This situation contrasts with the standard superkick scenario, where the atom is assumed to be much smaller than the photon \textcolor{black}{packet width} in order to probe approximately the same local momentum density of the photon coherently.

{ As highlighted in Refs. \cite{li2024unambiguous, liu2025superkick}, the superkick effect can  serve as a method for the unambiguous detection of a phase vortex. These technique can be particularly useful in experimental setups where conventional diagnostic tools — such as diffraction and interference — fail to work. The selfkick effect offers a similar pathway to probe vorticity in atomic and ionic wave packets. In this case, the atomic packet must expand to a transverse size of at \textit{least a few micrometers} so that its internal transverse momentum distribution can be resolved. This requires long free-expansion times and correspondingly low densities to suppress collisional decoherence; for ions, space-charge effects must be negligible over this scale. In addition, the photon beam must be tightly focused, with a transverse size $\lesssim 1 \mu$m. These simultaneous requirements make the implementation challenging but, in principle, achievable in setups employing nonparaxial photon beams in combination with sufficiently coherent trapped ions or atomic/ionic beams.}

\section{Discussion and conclusion}

We have theoretically investigated the interaction of nonrelativistic, spatially localized atomic packets with twisted photons in free space. Our framework treats both atom and photon as localized packets, goes beyond the dipole approximation, and fully incorporates atomic center-of-mass (CM) motion, enabling a complete quantum-mechanical description of absorption and scattering with recoil and non–plane-wave effects. { We predict line broadening due to finite coherence length of photonic wave packet, the superkick effect \cite{barnett2013superweak}, as well as it dual phenomenon - the selfkick, nonparaxial asymmetries and splitting of resonances caused by the photon’s longitudinal structure (the latter two effects are explained in Sec.~XII-XIII of Supplementary Materials).}

The central result is the transfer of orbital angular momentum (OAM) from a twisted photon to the atomic CM during absorption, enabled by the violation of standard selection rules. For dominant dipole transitions in head-on collisions, the most probable transferred OAM matches that of the photon. The resonant cross section can be extremely large (up to $10^9$ barn for hydrogen), suggesting a new route to generating twisted atoms complementary to diffraction \cite{luski2021vortex}. A promising experimental realization is a collinear atom–laser configuration \cite{Collinear_spec_1978, Collinear_spec_2021, baturin2024conversion}, for instance using beams produced by an Even--Lavie valve \cite{even2000cooling, luski2021vortex}.

Although our analysis focuses on free atoms, it readily extends to trapped ions, used in spectroscopy~\cite{schmiegelow2016transfer, schmiegelow2022, afanasev2018experimental}. While recoil is continuous for free ions and discrete in traps, the trapping energies ($10^{-10}\ldots10^{-8},\text{eV}$) are negligible compared with optical scales, so trapped ions effectively behave as free particles. This corresponds to replacing the continuous CM spectrum $\varepsilon^{\text{(CM)}}(\mathbf{P})$ with discrete levels and the coherence length with the trap length scale. The basis used here is closely related to cylindrical harmonic oscillator eigenfunctions \cite{mukherjee2018interaction, mondal2014angular, muthukrishnan2002entanglement, peshkov2023interaction}. More complex schemes can involve releasing trapped ions and illuminating them afterward \cite{fuhrmanek2010imaging, fuhrmanek2010measurement}.

Twisted atoms are of growing interest because their intrinsic OAM introduces new degrees of freedom in light–matter and matter–matter interactions, modifying selection rules and coupling internal and motional angular momentum. In scattering, their chirality can reveal observables inaccessible to plane-wave probes, analogous to twisted electrons and neutrons \cite{BLIOKH20171, afanasev2019schwinger, afanasev2021elastic, madan2020}. The generation of vortex ions for accelerator experiments is anticipated at the Institute of Modern Physics \cite{Huizhou}. In atom interferometry, CM OAM could enhance rotation sensitivity via $\ell$-dependent Sagnac-type phase shifts \cite{jaouadi2025towards}. While neutron interferometers have demonstrated related effects \cite{geerits2025measuring}, atomic systems offer narrower momentum spreads and greater control \cite{barrett2014sagnac, amico2021roadmap}.

Finally, vortex atoms can enable high-dimensional quantum information encoding. The quantized CM OAM provides an unbounded discrete degree of freedom suitable for qudits, extending concepts from photonic OAM encoding \cite{brandt2020high, kim2024qudit} and their storage in cold atoms \cite{nicolas2014quantum, ye2022long}. Motional states are already used as qubits in trapped ions \cite{bruzewicz2019trapped}, and extending this to OAM offers an angular-momentum basis that can be entangled with internal states \cite{muthukrishnan2002entanglement, jaouadi2025towards}.

\section{Acknowledgments}
We are grateful to Alexander Shchepkin for his assistance with the integrals and to Maxim Maximov for fruitful discussions. In addition, we warmly thank Andrey Surzhykov for many inspiring discussions and his valuable advice, which helped us a lot to improve our work. 

\section{Disclosures}
The authors declare no conflicts of interest.

\section{Data availability}
The data that support the findings of this study are available from the authors upon reasonable request.

\bibliographystyle{unsrt}
\bibliography{ref}

@article{BLIOKH20171,
title = {Theory and applications of free-electron vortex states},
journal = {Physics Reports},
volume = {690},
pages = {1-70},
year = {2017},
issn = {0370-1573},
doi = {https://doi.org/10.1016/j.physrep.2017.05.006},
url = {https://www.sciencedirect.com/science/article/pii/S0370157317301515},
author = {K.Y. Bliokh and I.P. Ivanov and G. Guzzinati and L. Clark and R. {Van Boxem} and A. B{\'e}ch{\'e} and R. Juchtmans and M.A. Alonso and P. Schattschneider and F. Nori and J. Verbeeck}
}

@article{NJP2021,
    author = "Karlovets, Dmitry",
    title = "{Vortex particles in axially symmetric fields and applications of the quantum Busch theorem}",
    doi = "10.1088/1367-2630/abeacc",
    journal = "New J. Phys.",
    volume = "23",
    number = "3",
    pages = "033048",
    year = "2021",
url = {https://dx.doi.org/10.1088/1367-2630/abeacc},
}

@article{schmiegelow2016transfer,
  title={Transfer of optical orbital angular momentum to a bound electron},
  author={Schmiegelow, Christian T and Schulz, Jonas and Kaufmann, Henning and Ruster, Thomas and Poschinger, Ulrich G and Schmidt-Kaler, Ferdinand},
  journal={Nature communications},
  volume={7},
  number={1},
  pages={12998},
  year={2016},
  publisher={Nature Publishing Group UK London}
}

@book{Mandel_Wolf_1995, place={Cambridge}, title={Optical Coherence and Quantum Optics}, publisher={Cambridge University Press}, author={Mandel, Leonard and Wolf, Emil}, year={1995}}

@article{scholz2014absorption,
  title={Absorption of twisted light by hydrogenlike atoms},
  author={Scholz-Marggraf, HM and Fritzsche, S and Serbo, VG and Afanasev, A and Surzhykov, A},
  journal={Physical Review A},
  volume={90},
  number={1},
  pages={013425},
  year={2014},
  publisher={APS}
}

@article{baturin2024conversion,
  title={Conversion of twistedness from light to atoms},
  author={Baturin, SS and Volotka, AV},
  journal={Physical Review A},
  volume={110},
  number={2},
  pages={L020801},
  year={2024},
  publisher={APS}
}

@article{EPJC2023,
  title={Shifting physics of vortex particles to higher energies via quantum entanglement},
  author={Karlovets, DV and Baturin, SS and Geloni, Gianluca and Sizykh, GK and Serbo, VG},
  journal={The European Physical Journal C},
  volume={83},
  number={5},
  pages={372},
  year={2023},
  publisher={Springer}
}

@article{EPJC2024,
  title={Generation of vortex particles via generalized measurements},
  author={Karlovets, DV and Baturin, SS and Geloni, G and Sizykh, GK and Serbo, VG},
  journal={The European Physical Journal C},
  volume={82},
  number={11},
  pages={1008},
  year={2022},
  publisher={Springer}
}

@article{Pavlov2024,
  title = {Generation of vortex electrons by atomic photoionization},
  author = {Pavlov, I. I. and Chaikovskaia, A. D. and Karlovets, D. V.},
  journal = {Phys. Rev. A},
  volume = {110},
  issue = {3},
  pages = {L031101},
  numpages = {7},
  year = {2024},
  month = {Sep},
  publisher = {American Physical Society},
  doi = {10.1103/PhysRevA.110.L031101},
  url = {https://link.aps.org/doi/10.1103/PhysRevA.110.L031101}
}

@article{maslennikov2024theoretical,
  title={Theoretical consideration of a twisted atom},
  author={Maslennikov, PK and Volotka, AV and Baturin, SS},
  journal={Physical Review A},
  volume={109},
  number={5},
  pages={052805},
  year={2024},
  publisher={APS}
}

@article{ivanov2022double,
  title={Double-Twisted Spectroscopy with Delocalized Atoms},
  author={Ivanov, Igor P},
  journal={Annalen der Physik},
  volume={534},
  number={3},
  pages={2100128},
  year={2022},
  publisher={Wiley Online Library}
}

@article{luski2021vortex,
  title={Vortex beams of atoms and molecules},
  author={Luski, Alon and Segev, Yair and David, Rea and Bitton, Ora and Nadler, Hila and Barnea, A Ronny and Gorlach, Alexey and Cheshnovsky, Ori and Kaminer, Ido and Narevicius, Edvardas},
  journal={Science},
  volume={373},
  number={6559},
  pages={1105--1109},
  year={2021},
  publisher={American Association for the Advancement of Science}
}

@article{ivanov2022promises,
  title={Promises and challenges of high-energy vortex states collisions},
  author={Ivanov, Igor P},
  journal={Progress in Particle and Nuclear Physics},
  volume={127},
  pages={103987},
  year={2022},
  publisher={Elsevier}
}

@article{forbes2021structured,
  title={Structured light},
  author={Forbes, Andrew and De Oliveira, Michael and Dennis, Mark R},
  journal={Nature Photonics},
  volume={15},
  number={4},
  pages={253--262},
  year={2021},
  publisher={Nature Publishing Group UK London}
}

@article{roadmap_strwaves,
  title={Roadmap on structured waves},
  author={Bliokh, Konstantin Y and Karimi, Ebrahim and Padgett, Miles J and Alonso, Miguel A and Dennis, Mark R and Dudley, Angela and Forbes, Andrew and Zahedpour, Sina and Hancock, Scott W and Milchberg, Howard M and others},
  journal={Journal of Optics},
  volume={25},
  number={10},
  pages={103001},
  year={2023},
  publisher={IOP Publishing}
}

@article{roadmap_strlight,
  title={Roadmap on structured light},
  author={Rubinsztein-Dunlop, Halina and Forbes, Andrew and Berry, Michael V and Dennis, Mark R and Andrews, David L and Mansuripur, Masud and Denz, Cornelia and Alpmann, Christina and Banzer, Peter and Bauer, Thomas and others},
  journal={Journal of Optics},
  volume={19},
  number={1},
  pages={013001},
  year={2016},
  publisher={IOP Publishing}
}

@article{knyazev2018beams,
  title={Beams of photons with nonzero projections of orbital angular momenta: New results},
  author={Knyazev, Boris Aleksandrovich and Serbo, VG},
  journal={Physics-Uspekhi},
  volume={61},
  number={5},
  pages={449},
  year={2018},
  publisher={IOP Publishing}
}

@article{JHEP2017,
  title={Scattering of wave packets with phases},
  author={Karlovets, Dmitry V},
  journal={Journal of High Energy Physics},
  volume={2017},
  number={3},
  pages={1--46},
  year={2017},
  publisher={Springer}
}

@article{karlovets2020effects,
  title={Effects of the transverse coherence length in relativistic collisions},
  author={Karlovets, Dmitry V and Serbo, Valeriy G},
  journal={Physical Review D},
  volume={101},
  number={7},
  pages={076009},
  year={2020},
  publisher={APS}
}

@article{liu2023threshold,
  title={Threshold effects in high-energy vortex state collisions},
  author={Liu, Bei and Ivanov, Igor P},
  journal={Physical Review A},
  volume={107},
  number={6},
  pages={063110},
  year={2023},
  publisher={APS}
}

@article{fedorov2005spontaneous,
  title={Spontaneous emission of a photon: Wave-packet structures and atom-photon entanglement},
  author={Fedorov, MV and Efremov, MA and Kazakov, AE and Chan, KW and Law, CK and Eberly, JH},
  journal={Physical Review A—Atomic, Molecular, and Optical Physics},
  volume={72},
  number={3},
  pages={032110},
  year={2005},
  publisher={APS}
}

@article{karlovets2018relativistic,
  title={Relativistic vortex electrons: Paraxial versus nonparaxial regimes},
  author={Karlovets, Dmitry},
  journal={Physical Review A},
  volume={98},
  number={1},
  pages={012137},
  year={2018},
  publisher={APS}
}

@book{messiah2014quantum,
  title={Quantum mechanics},
  author={Messiah, Albert},
  year={2014},
  publisher={Courier Corporation}
}

@book{sakurai1967advanced,
  title={Advanced Quantum Mechanics},
  author={Sakurai, J.J.},
  isbn={9780201067101},
  lccn={67019430},
  series={A-W series in advanced physics},
  url={https://books.google.ru/books?id=ZXEsAAAAYAAJ},
  year={1967},
  publisher={Addison-Wesley Publishing Company}
}

@article{kotkin1992processes,
  title={Processes with large impact parameters at colliding beams},
  author={Kotkin, GL and Serbo, VG and Schiller, A},
  journal={International Journal of Modern Physics A},
  volume={7},
  number={20},
  pages={4707--4745},
  year={1992},
  publisher={World Scientific}
}

@article{chaikovskaia2024vavilov,
  title={Vavilov-Cherenkov emission with twisted electrons: A study of the final entangled state},
  author={Chaikovskaia, Alisa D and Karlovets, Dmitry V and Serbo, VG},
  journal={Physical Review A},
  volume={109},
  number={1},
  pages={012222},
  year={2024},
  publisher={APS}
}

@article{ivanov2011colliding,
  title={Colliding particles carrying nonzero orbital angular momentum},
  author={Ivanov, Igor P},
  journal={Physical Review D—Particles, Fields, Gravitation, and Cosmology},
  volume={83},
  number={9},
  pages={093001},
  year={2011},
  publisher={APS}
}

@article{karlovets2025attosecond,
  title={Attosecond physics hidden in Cherenkov radiation},
  author={Karlovets, Dmitry and Chaikovskaia, Alisa and Grosman, Dmitriy and Kargina, Daria and Shchepkin, Aleksandr and Sizykh, Georgii},
  journal={Communications Physics},
  volume={8},
  number={1},
  pages={192},
  year={2025},
  publisher={Nature Publishing Group UK London}
}

@book{Varshalovich,
  title={Quantum Theory of Angular Momentum},
  author={Dmitrii A. Varshalovich and A. N. Moskalev and V. K. Khersonskii},
  year={1988},
  url={https://api.semanticscholar.org/CorpusID:117798939}
}

@book{BLP,
  title={Quantum Electrodynamics: Volume 4},
  author={Berestetskii, Vladimir Borisovich and Lifshitz, Evgenii Mikhailovich and Pitaevskii, Lev Petrovich},
  volume={4},
  year={1982},
  publisher={Butterworth-Heinemann}
}

@article{melezhik2025acceleration,
  title={Acceleration and twisting of neutral atoms by strong elliptically polarized short-wavelength laser pulses},
  author={Melezhik, Vladimir S and Shadmehri, Sara},
  journal={The Journal of Chemical Physics},
  volume={162},
  number={17},
  year={2025},
  publisher={AIP Publishing}
}

@article{even2000cooling,
  title={Cooling of large molecules below 1 K and He clusters formation},
  author={Even, Uzi and Jortner, J and Noy, D and Lavie, N and Cossart-Magos, C},
  journal={The Journal of Chemical Physics},
  volume={112},
  number={18},
  pages={8068--8071},
  year={2000},
  publisher={American Institute of Physics}
}

@article{schmiegelow2022,
  title={Coherent transfer of transverse optical momentum to the motion of a single trapped ion},
  author={Stopp, Felix and Verde, Maurizio and Katz, Milton and Drechsler, Martin and Schmiegelow, Christian T and Schmidt-Kaler, Ferdinand},
  journal={Physical Review Letters},
  volume={129},
  number={26},
  pages={263603},
  year={2022},
  publisher={APS}
}

@article{bruzewicz2019trapped,
  title={Trapped-ion quantum computing: Progress and challenges},
  author={Bruzewicz, Colin D and Chiaverini, John and McConnell, Robert and Sage, Jeremy M},
  journal={Applied physics reviews},
  volume={6},
  number={2},
  year={2019},
  publisher={AIP Publishing}
}

@article{Collinear_spec_1978,
  title = {Collinear Laser Spectroscopy on Fast Atomic Beams},
  author = {Anton, K. -R. and Kaufman, S. L. and Klempt, W. and Moruzzi, G. and Neugart, R. and Otten, E. -W. and Schinzler, B.},
  journal = {Phys. Rev. Lett.},
  volume = {40},
  issue = {10},
  pages = {642--645},
  numpages = {0},
  year = {1978},
  month = {Mar},
  publisher = {American Physical Society},
  doi = {10.1103/PhysRevLett.40.642},
  url = {https://link.aps.org/doi/10.1103/PhysRevLett.40.642}
}

@article{Collinear_spec_2021,
  title = {Beam energy determination via collinear laser spectroscopy},
  author = {K\"onig, Kristian and Minamisono, Kei and Lantis, Jeremy and Pineda, Skyy and Powel, Robert},
  journal = {Phys. Rev. A},
  volume = {103},
  issue = {3},
  pages = {032806},
  numpages = {7},
  year = {2021},
  month = {Mar},
  publisher = {American Physical Society},
  doi = {10.1103/PhysRevA.103.032806},
  url = {https://link.aps.org/doi/10.1103/PhysRevA.103.032806}
}

@article{fuhrmanek2010imaging,
  title={Imaging a single atom in a time-of-flight experiment},
  author={Fuhrmanek, Andreas and Lance, Andrew Matheson and Tuchendler, Charles and Grangier, Philippe and Sortais, Yvan RP and Browaeys, Antoine},
  journal={New Journal of Physics},
  volume={12},
  number={5},
  pages={053028},
  year={2010},
  publisher={IOP Publishing}
}

@article{fuhrmanek2010measurement,
  title={Measurement of the atom number distribution in an optical tweezer using single-photon counting},
  author={Fuhrmanek, Andreas and Sortais, Yvan RP and Grangier, Philippe and Browaeys, Antoine},
  journal={Physical Review A—Atomic, Molecular, and Optical Physics},
  volume={82},
  number={2},
  pages={023623},
  year={2010},
  publisher={APS}
}

@article{afanasev2021elastic,
  title={Elastic scattering of twisted neutrons by nuclei},
  author={Afanasev, AV and Karlovets, DV and Serbo, VG},
  journal={Physical Review C},
  volume={103},
  number={5},
  pages={054612},
  year={2021},
  publisher={APS}
}

@article{afanasev2019schwinger,
  title={Schwinger scattering of twisted neutrons by nuclei},
  author={Afanasev, Andrei V and Karlovets, DV and Serbo, VG},
  journal={Physical Review C},
  volume={100},
  number={5},
  pages={051601},
  year={2019},
  publisher={APS}
}

@article{hayrapetyan2013bessel,
  title={Bessel beams of two-level atoms driven by a linearly polarized laser field},
  author={Hayrapetyan, Armen G and Matula, Oliver and Surzhykov, Andrey and Fritzsche, Stephan},
  journal={The European Physical Journal D},
  volume={67},
  number={8},
  pages={167},
  year={2013},
  publisher={Springer}
}

@article{lembessis2014atom,
  title={Atom vortex beams},
  author={Lembessis, VE and Ellinas, D and Babiker, Mohamed and Al-Dossary, O},
  journal={Physical Review A},
  volume={89},
  number={5},
  pages={053616},
  year={2014},
  publisher={APS}
}

@article{helseth2004atomic,
  title={Atomic vortex beams in focal regions},
  author={Helseth, LE},
  journal={Physical Review A},
  volume={69},
  number={1},
  pages={015601},
  year={2004},
  publisher={APS}
}

@book{Scully,
  title={Quantum optics},
  author={ Scully,M.O. and Zubairy,M.S. },
  year={1997},
  publisher={Cambridge university press}
}

@article{liu2026vortex,
  title={Vortex-photon-induced multipole transitions in atomic nuclei},
  author={Liu, Shiyu and Ji, Liangliang},
  journal={Physical Review A},
  volume={113},
  number={1},
  pages={013121},
  year={2026},
  publisher={APS}
}

@article{barrett2014sagnac,
  title={The Sagnac effect: 20 years of development in matter-wave interferometry},
  author={Barrett, Brynle and Geiger, R{\'e}my and Dutta, Indranil and Meunier, Matthieu and Canuel, Benjamin and Gauguet, Alexandre and Bouyer, Philippe and Landragin, Arnaud},
  journal={Comptes Rendus. Physique},
  volume={15},
  number={10},
  pages={875--883},
  year={2014}
}

@article{geerits2025measuring,
  title={Measuring the angular momentum of a neutron using Earth's rotation},
  author={Geerits, Niels and Sponar, Stephan and Steffen, Kyle E and Snow, William M and Parnell, Steven R and Mauri, Giacomo and Smith, Gregory N and Dalgliesh, Robert M and de Haan, Victor},
  journal={Physical Review Research},
  volume={7},
  number={1},
  pages={013046},
  year={2025},
  publisher={APS}
}

@article{amico2021roadmap,
  title={Roadmap on Atomtronics: State of the art and perspective},
  author={Amico, Luigi and Boshier, Malcolm and Birkl, Gerhard and Minguzzi, Anna and Miniatura, Christian and Kwek, L-C and Aghamalyan, Davit and Ahufinger, Veronica and Anderson, Dana and Andrei, Natan and others},
  journal={AVS Quantum Science},
  volume={3},
  number={3},
  year={2021},
  publisher={AIP Publishing}
}

@article{nicolas2014quantum,
  title={A quantum memory for orbital angular momentum photonic qubits},
  author={Nicolas, A and Veissier, L and Giner, L and Giacobino, E and Maxein, D and Laurat, J},
  journal={Nature Photonics},
  volume={8},
  number={3},
  pages={234--238},
  year={2014},
  publisher={Nature Publishing Group UK London}
}

@article{ye2022long,
  title={Long-lived memory for orbital angular momentum quantum states},
  author={Ye, Ying-Hao and Zeng, Lei and Dong, Ming-Xin and Zhang, Wei-Hang and Li, En-Ze and Li, Da-Chuang and Guo, Guang-Can and Ding, Dong-Sheng and Shi, Bao-Sen},
  journal={Physical Review Letters},
  volume={129},
  number={19},
  pages={193601},
  year={2022},
  publisher={APS}
}

@article{brandt2020high,
  title={High-dimensional quantum gates using full-field spatial modes of photons},
  author={Brandt, Florian and Hiekkam{\"a}ki, Markus and Bouchard, Fr{\'e}d{\'e}ric and Huber, Marcus and Fickler, Robert},
  journal={Optica},
  volume={7},
  number={2},
  pages={98--107},
  year={2020},
  publisher={Optical Society of America}
}

@article{kim2024qudit,
  title={Qudit-based variational quantum eigensolver using photonic orbital angular momentum states},
  author={Kim, Byungjoo and Hu, Kang-Min and Sohn, Myung-Hyun and Kim, Yosep and Kim, Yong-Su and Lee, Seung-Woo and Lim, Hyang-Tag},
  journal={Science Advances},
  volume={10},
  number={43},
  pages={eado3472},
  year={2024},
  publisher={American Association for the Advancement of Science}
}

@article{mondal2014angular,
  title={Angular momentum transfer in interaction of Laguerre-Gaussian beams with atoms and molecules},
  author={Mondal, Pradip Kumar and Deb, Bimalendu and Majumder, Sonjoy},
  journal={Physical Review A},
  volume={89},
  number={6},
  pages={063418},
  year={2014},
  publisher={APS}
}

@article{mukherjee2018interaction,
  title={Interaction of a Laguerre--Gaussian beam with trapped Rydberg atoms},
  author={Mukherjee, Koushik and Majumder, Sonjoy and Mondal, Pradip Kumar and Deb, Bimalendu},
  journal={Journal of Physics B: Atomic, Molecular and Optical Physics},
  volume={51},
  number={1},
  pages={015004},
  year={2018},
  publisher={IOP Publishing}
}

@article{muthukrishnan2002entanglement,
  title={Entanglement of internal and external angular momenta of a single atom},
  author={Muthukrishnan, Ashok and Stroud Jr, CR},
  journal={Journal of Optics B: Quantum and Semiclassical Optics},
  volume={4},
  number={2},
  pages={S73--S77},
  year={2002}
}

@article{peshkov2023interaction,
  title={Interaction of twisted light with a trapped atom: Interplay between electronic and motional degrees of freedom},
  author={Peshkov, AA and Bidasyuk, YM and Lange, R and Huntemann, N and Peik, E and Surzhykov, A},
  journal={Physical Review A},
  volume={107},
  number={2},
  pages={023106},
  year={2023},
  publisher={APS}
}

@article{barnett2013superweak,
  title={Superweak momentum transfer near optical vortices},
  author={Barnett, Stephen M and Berry, MV},
  journal={Journal of Optics},
  volume={15},
  number={12},
  pages={125701},
  year={2013},
  publisher={IOP Publishing}
}

@article{ivanov2022observability,
  title={Observability of the superkick effect within a quantum-field-theoretical approach},
  author={Ivanov, Igor P and Liu, Bei and Zhang, Pengming},
  journal={Physical Review A},
  volume={105},
  number={1},
  pages={013522},
  year={2022},
  publisher={APS}
}

@article{li2024unambiguous,
  title={Unambiguous detection of high-energy vortex states via the superkick effect},
  author={Li, Zhengjiang and Liu, Shiyu and Liu, Bei and Ji, Liangliang and Ivanov, Igor P},
  journal={Physical Review Letters},
  volume={133},
  number={26},
  pages={265001},
  year={2024},
  publisher={APS}
}

@article{afanasev2021recoil,
  title={Recoil momentum effects in quantum processes induced by twisted photons},
  author={Afanasev, Andrei and Carlson, Carl E and Mukherjee, Asmita},
  journal={Physical Review Research},
  volume={3},
  number={2},
  pages={023097},
  year={2021},
  publisher={APS}
}

@article{afanasev2022superkicks,
  title={Superkicks and the photon angular and linear momentum density},
  author={Afanasev, Andrei and Carlson, Carl E and Mukherjee, Asmita},
  journal={Physical Review A},
  volume={105},
  number={6},
  pages={L061503},
  year={2022},
  publisher={APS}
}

@article{liu2025superkick,
  title={Superkick effect in vortex scattering: M{\o}ller scattering},
  author={Liu, Shiyu and Liu, Bei and Ivanov, Igor P and Ji, Liangliang},
  journal={Physical Review A},
  volume={112},
  number={3},
  pages={032814},
  year={2025},
  publisher={APS}
}

@article{Huizhou,
  title={High-precision physics experiments at Huizhou large-scale scientific facilities},
  author={An, Fengpeng and Bai, Dong and Cai, Hanjie and Chen, Siyuan and Chen, Xurong and Duyang, Hongyue and Gao, Leyun and Ge, Shaofeng and He, Jun and Huang, Junting and others},
  journal={Chinese Physics Letters},
  volume={42},
  number={11},
  pages={110102},
  year={2025},
  publisher={Chinese Physical Society and IOP Publishing Ltd}
}

@article{shchepkin2025absorption,
  title={Absorption of a twisted photon by an electron in strong magnetic field},
  author={Shchepkin, AA and Grosman, DV and Shkarupa, II and Karlovets, DV},
  journal={The European Physical Journal C},
  volume={85},
  number={1},
  pages={11},
  year={2025},
  publisher={Springer}
}

@article{lembessis2017atomic,
  title={Atomic Ferris wheel beams},
  author={Lembessis, Vasileios E},
  journal={Physical Review A},
  volume={96},
  number={1},
  pages={013622},
  year={2017},
  publisher={APS}
}

@article{madan2020,
  title={The quantum future of microscopy: Wave function engineering of electrons, ions, and nuclei},
  author={Madan, Ivan and Vanacore, Giovanni Maria and Gargiulo, Simone and LaGrange, Thomas and Carbone, Fabrizio},
  journal={Applied Physics Letters},
  volume={116},
  number={23},
  year={2020},
  publisher={AIP Publishing}
}

@article{afanasev2018experimental,
  title={Experimental verification of position-dependent angular-momentum selection rules for absorption of twisted light by a bound electron},
  author={Afanasev, Andrei and Carlson, Carl E and Schmiegelow, Christian T and Schulz, Jonas and Schmidt-Kaler, Ferdinand and Solyanik, Maria},
  journal={New Journal of Physics},
  volume={20},
  number={2},
  pages={023032},
  year={2018},
  publisher={IOP Publishing}
}

@article{alharbi2023photoexcitation,
  title={Photoexcitation of atoms near the center of vortex light},
  author={Alharbi, Abdullah F and Lyras, A and Lembessis, Vassilis E and Al-Dossary, Omar},
  journal={Results in Physics},
  volume={46},
  pages={106311},
  year={2023},
  publisher={Elsevier}
}

@inproceedings{jaouadi2025towards,
  title={Towards a Twisted Atom Laser: Cold Atoms Released from Helical Optical Tube Potentials},
  author={Jaouadi, Amine and Lyras, Andreas and Lembessis, Vasileios E},
  booktitle={Photonics},
  volume={12},
  number={10},
  pages={999},
  year={2025},
  organization={MDPI}
}

@article{zanon2023engineering,
  title={Engineering quantum control with optical transitions induced by twisted light fields},
  author={Zanon-Willette, Thomas and Impens, Fran{\c{c}}ois and Arimondo, Ennio and Wilkowski, David and Taichenachev, Aleksei V and Yudin, Valeriy I},
  journal={Physical Review A},
  volume={108},
  number={4},
  pages={043513},
  year={2023},
  publisher={APS}
}

@article{babiker2019atoms,
  title={Atoms in complex twisted light},
  author={Babiker, Mohamed and Andrews, David L and Lembessis, Vassilis E},
  journal={Journal of Optics},
  volume={21},
  number={1},
  pages={013001},
  year={2019},
  publisher={IOP Publishing}
}

\end{document}


\title{Supplemental Material to ``Interaction of twisted light with free twisted atoms''}
\author{I.~Pavlov}
\author{A.~Chaikovskaia}
\author{D.~Karlovets}

\date{\today}

\maketitle

\onecolumngrid
\section{Free packet of a massive particle}\label{FrShr}

Consider a non-relativistic freely propagating wave packet of a charged particle with a mass $\mu$ described with the Schr\"odinger equation -- be it an electron, a proton, or an ion. We start with a one-dimensional (1D) case, for which there is an exact non-stationary solution, a generalization of the gaussian packet, called {\it a standard Hermite-gaussian (HG) packet} with an integer quantum number $k=0,1,2,...$,\cite{NJP2021},
\bea
\label{HG_pos}
& \displaystyle  \psi_{||}^{(k)}(z,t) = \frac{i^k}{\sqrt{2^k k! \sqrt{\pi} \sigma_z(t)}}\, H_k\left(\frac{z - \la u\ra t}{\sigma_z(t)}\right) \exp\Big\{-it\frac{\la p\ra^2}{2\mu} + i \la p\ra z -\cr
& \displaystyle -i \left(k+\fr{1}{2}\right) \arctan\frac{t}{t_{d,z}} - \frac{(z - \la u\ra t)^2}{2\sigma_z^2(t)} \left(1-i\frac{t}{t_{d,z}}\right)\Big\},\cr
& \displaystyle \int\limits_{-\infty}^{+\infty} dz\, |\psi_{||}^{(k)}(z,t)|^2 = 1,
\label{HGpsi}
\eea
where the term with $\arctan t/t_{d,z}$ is called the Gouy phase,
\bea
\la u\ra = \fr{\la p\ra}{\mu} = \frac{1}{\mu}\int\limits_{-\infty}^{+\infty} dz\,\left(\psi_{||}^{(k)}(z,t)\right)^*\hat{p}\,\psi_{||}^{(k)}(z,t)
\eea
is the mean velocity, 
$$
t_{d,z} = \mu\, \sigma_z^2(0) \equiv \frac{\sigma_z^2(0)}{c \lambda_c}
$$ 
is the diffraction time, $\lambda_c = 1/\mu \equiv/\mu c$ is the Compton wavelength. Here also
$$
\sigma_z(t) = \sigma_z(0) \sqrt{1 + t^2/t_{d,z}^2},
$$
and one employs the orthogonality relation, 
\bea
\int\limits_{-\infty}^{+\infty} dx\, H_n(x) H_m(x) e^{-x^2} = 2^n n! \sqrt{\pi}\, \delta_{nm}.
\eea

This HG packet can be obtained starting from the momentum representation, in which
\bea
\label{HG_mom}
& \displaystyle  \psi_{||}^{(k)}(p,t) = \sqrt{\fr{2\pi\sigma_z}{2^k k! \sqrt{\pi}}}\, H_k\left(\sigma_z\left(p-\la p\ra\right)\right) \exp\Big\{-it\frac{p^2}{2\mu} - \fr{\sigma_z^2}{2}\left(p-\la p\ra\right)^2\Big\},\cr
& \displaystyle \int\limits_{-\infty}^{+\infty} \fr{dp}{2\pi}\, |\psi_{||}^{(k)}(p,t)|^2 = 1.
\label{HGpsiP}
\eea
One can estimate the diffraction time
\bea
t_{d,z} = \left(\frac{\sigma_z}{c}\right)^2\frac{1}{t_c},\ t_c = \lambda_c/c,
\eea
where $t_c \approx 1.29\times 10^{-21}$ s for electron and $t_c \approx 0.7\times 10^{-24}$ s for proton. The packets of atoms and ions in traps can nowadays be localized with the precision of $\sigma_z \sim 10$ nm \cite{schmiegelow2016transfer}, which yields
\bea
t_{d,z}^{(p)} \approx 1.6\, \text{ns}
\label{Ns}
\eea
for proton and some $16$ ps for $\sigma_z \sim 1$ nm. As the diffraction time grows with the particle mass, $t_{d,z}^{(p)} \propto \mu$, the former is even larger for heavy atoms.

Let us assume that the packet moves in a straight line in 3-dimensional (3D) space, 
\be
\la z\ra = \la u\ra t, 
\ee
and that the longitudinal dynamics is decoupled from the transverse one. In the transverse plane, the packet can be described as {\it a standard Laguerre-gaussian beam} with a quantized angular momentum projection on the propagation direction $\ell = 0, \pm 1, \pm 2, ...$ and the principal quantum number $n=0,1,2,...$,
\bea
\label{LG_pos}
& \displaystyle \psi_{\perp}^{(n,\ell)}({\bm \rho},t) = \sqrt{\frac{n!}{(n + |\ell|)!}}\, \frac{1}{\sqrt{\pi}}\frac{\rho^{|\ell|}}{\sigma_{\perp}^{|\ell|+1}(t)}\, L_{n}^{|\ell|}\left(\frac{\rho^2}{\sigma_{\perp}^2(t)}\right) \exp\Big\{i\ell \phi - i (2n + |\ell| + 1)\arctan\fr{t}{t_{d,\perp}} - \cr 
& \displaystyle - \frac{\rho^2}{2\sigma_{\perp}^2(t)}\, \left(1-i\fr{t}{t_{d,\perp}}\right)\Big\},\ \int d^2 {\bm \rho}\, |\psi_{\perp}^{(n,\ell)}({\bm \rho},t)|^2 = 1,
\label{LGpsi}
\eea
where $t_{d,\perp} = \mu\, \sigma_{\perp}^2(0)\equiv \sigma_{\perp}^2(0)/(c \lambda_c), \sigma_{\perp}(t) = \sigma_{\perp}(0) \sqrt{1 + t^2/t_{d,\perp}^2}$. The wave function of the packet localized in 3D space looks as
\bea
\label{HG_LG}
\displaystyle & \psi^{(k, n,\ell)}(\br,t) = \psi_{||}^{(k)}(z,t) \psi_{\perp}^{(n,\ell)}({\bm \rho},t),\cr
\displaystyle & \int d^3x\, |\psi^{(k, n,\ell)}(\br,t)|^2 = 1. 
\eea
The set of these solutions, which we will call {\it an HG$\times$LG packet}, is parameterized with three quantum numbers $k,n,\ell$;
it is complete and orthogonal, and $\psi^{(k, n,\ell)}(\br,t)$ exactly satisfies the free non-stationary Schr\"odinger equation.

The integers at the Gouy phases, 
\bea
\displaystyle & k+1/2\ \text{in 1D},\cr
\displaystyle & 2n+|\ell|+1\ \text{in 2D}, 
\eea
are called {\it the quality factors} of the packet and they are measures of its ``quantumness'' because we get a gaussian state for $n=\ell=k=0$. Non-classicality of the state could also be expressed through the concept of {\it quantum emittance}, defined as
\begin{equation}
\epsilon_{z} \equiv \sqrt{\langle z^2\rangle \langle {u}^2\rangle - \langle z {u} \rangle \langle {u} z\rangle },
\end{equation}
\begin{equation}
\label{Emitfh}
    \epsilon_i = \sqrt{\langle x_i^2\rangle \langle \hat{u}_i^2\rangle - \langle x_i \hat{u}_i\rangle \langle \hat{u}_i x_i \rangle } = \frac{1}{2} \sqrt{\langle \bm{\rho}^2\rangle \langle \hat{\bm{u}}_{\perp}^2\rangle - \langle \bm{\rho} \cdot \hat{\bm{u}}_{\perp}\rangle \langle \hat{\bm{u}}_{\perp} \cdot \bm{\rho} \rangle } \equiv \frac{\epsilon_{\bot}}{2},
\end{equation}
where $i$ enumerates the two transverse axes. For the above example we find
\begin{align}
\label{em_z}
\epsilon_{z} &= \lambda_{c}\left(k+\frac12 \right), \\ 
\label{em_tr}
    \epsilon_{\bot} &= \lambda_{c}\sqrt{(2n+|l|+1)^2 - 1}.
\end{align}
Thus, the minimal transverse emittance is vanishing, $\epsilon_{\bot}(n=l=0)=0$, whereas the longitudinal one is not, $\epsilon_{z}(k=0) = \lambda_c/2$. The quality factors can be expressed via these emittances as
\bea
& k+1/2 = \epsilon_{z}/\lambda_c,\\
& 2n+|l|+1 = \sqrt{\left(\epsilon_{\bot}/\lambda_c\right)^2 + 1}.
\eea

{It is instructive to calculate the wave function in momentum representation,
\bea
\label{LG_mom}
\displaystyle \psi_{\perp}^{(n,\ell)}(\bp_{\perp},t) = \int d{\bm \rho}\,\psi_{\perp}^{(n,\ell)}({\bm \rho},t)\,e^{-i\bp_{\perp}\cdot{\bm \rho}};\\
\displaystyle \psi_{\perp}^{(n,\ell)}(\bp_{\perp},0) = (-1)^n i^{-\ell} 2 \sqrt{\frac{\pi n!}{(n + |\ell|)!}}  \sigma_{\perp}^{\ell + 1} p_\perp^\ell \exp\left\{-\frac{p_\perp^2\sigma_{\perp}^2}{2}\right\} L_n^\ell\left({p_\perp^2 \sigma_{\perp}^2}\right)\exp\left\{i\ell \phi_p\right\},\\ \int\fr{d\bp_{\perp}}{(2\pi)^2}\,\left|\psi_{\perp}^{(n,\ell)}(\bp_{\perp},t)\right|^2=1.
\eea
For the azimuthal integral, let us substitute $\bp_{\perp}\cdot{\bm \rho}= p_{\perp}\rho \cos(\phi-\phi_p)$, $d{\bm \rho}=\rho d\rho d\phi$. Calculating the azimuthal integral is straightforward with the use of Jacobi–Anger expansion 
\be
\int d \phi e^{i\ell \phi - i p_{\perp}\rho \cos(\phi-\phi_p)}=(-i)^\ell 2\pi J_\ell( p_{\perp}\rho)e^{i\ell \phi_p}.
\ee
}

\section{A wave packet of a free hydrogen atom}\label{HH}

Consider now the time-dependent Schr\"odinger equation for a free Hydrogen atom:
\bea
i\partial_t \Psi(\br_e,\br_n,t) = \left(\frac{\hat{\bp}^2_e}{2m} + \frac{\hat{\bp}^2_n}{2M} + U(r)\right)\Psi(\br_e,\br_n,t)
\eea
where $r = |\br_e-\br_n|$, $U(r) = -e^2/r$. Going to the following variables
\bea
\br = \br_e - \br_n,\ \bR = \frac{m\br_e + M\br_n}{m+M},\ \mu = \frac{m M}{m + M},
\label{rR}
\eea
we arrive at
\bea
& \displaystyle i\partial_t \Psi(\br,\bR,t) = 
\underbrace{\left(\frac{\hat{\bp}^2_r}{2\mu} + \frac{\hat{\bp}^2_R}{2(m + M)} + U(r)\right)}_{\hat{H}^{(at)}}\Psi(\br,\bR,t).
\eea
Now we split the wave function as 
\bea
\Psi(\br,\bR,t) = \psi(\br,t) \phi(\bR,t),
\eea
both new functions being time dependent. For the electron component we come to the \textit{stationary} Schr\"odinger equation
\bea
& \displaystyle \psi(\br,t) = \psi(\br)\, \exp\left\{-i\varepsilon_e t\right\},\cr
& \displaystyle \varepsilon_e\, \psi(\br) = \left(\frac{\hat{\bp}^2_r}{2\mu} + U(r)\right)\psi(\br),
\eea
so the electronic degrees of freedom and the center-of-mass motion decouple.

For the center-of-mass wave function, we find the \textit{non-stationary} free Schr\"odinger equation
\bea
& \displaystyle i\partial_t \phi(\bR,t) = \frac{\hat{\bp}^2_R}{2(m + M)}\,\phi(\bR,t).
\eea
Along with a plane wave, {\it its exact solution} is a product (\ref{HG_LG}) of the LG packet in the transverse plane (\ref{LGpsi}) and the HG one (\ref{HGpsi}) in the longitudinal plane with $\mu \to m+M$:
\bea
& \displaystyle \phi(\bR,t) \equiv \phi_{k,n,\ell}(\bR,t) = \phi_{||}^{(k)}(z,t) \phi_{\perp}^{(n,\ell)}(\bR,t).
\eea
The packet of the center-of-mass moves on average in a straight line, 
\bea
& \displaystyle \la z\ra = \frac{\la p\ra}{m+M}\, t \equiv \la u\ra_{cm} t,
\eea
and it spreads in the transverse plane with time according to 
\bea
& \displaystyle \la\rho^2\ra(t) = \la\rho^2\ra(0) + \frac{\partial \la\rho^2\ra(0)}{\partial t}\, t + \la{\bm u}_{\perp}^2\ra\, t^2,
\eea
which follows from the Heisenberg equations for the $\rho^2$ operator \cite{NJP2021}. If the focus of the packet is in the initial moment of time, i.e. $\partial \la\rho^2\ra(0)/\partial t=0$, then 
\bea
& \displaystyle \la\rho^2\ra(t) = \la\rho^2\ra(0) + \la{\bm u}_{\perp}^2\ra\, t^2 = \la\rho^2\ra(0) \left(1 + \frac{t^2}{t_{d,\perp}^2}\right) = \la\rho^2\ra(0) \left(1 + \frac{\la z\ra^2}{z_R^2}\right),
\eea
where $t_{d,\perp} = \sqrt{\la\rho^2\ra(0)/\la{\bm u}_{\perp}^2\ra}$ is the diffraction time and $z_R = \la u\ra_{cm} t_{d,\perp}$ is \textit{the Reyleigh length}. When spreading is essential, $\la z\ra \gg z_R$, we have 
\bea
& \displaystyle \rho(\la z\ra) \equiv \sqrt{\la\rho^2\ra(\la z\ra)} \approx \sqrt{\la{\bm u}_{\perp}^2\ra}\, t,
\eea

Within the model of the LG packet (\ref{HG_LG}), one can calculate the averages and arrive at\footnote{Note the misprint in equations (76) and (111) of Ref.\cite{NJP2021}: instead of $n + |\ell| + 1$ there must be $2n + |\ell| + 1$.} \cite{NJP2021}
\bea 
& \displaystyle \sqrt{\la {\bm u}_{\perp}^2\ra} = \frac{1}{(m+M)\, \sigma_{\perp}(0)}\, \sqrt{2n + |\ell| + 1},\cr
& \displaystyle \sqrt{\la \rho^2\ra(0)} = \sigma_{\perp}(0)\, \sqrt{2n + |\ell| + 1},\cr
& \displaystyle \text{and so}\quad t_{d,\perp} = \fr{\sqrt{\la\rho^2\ra(0)}}{\sqrt{\la{\bm u}_{\perp}^2\ra}} = (m+M)\sigma_{\perp}^2(0),
\label{rmsm}
\eea
in accord with Eq. \eqref{LGpsi}. If we introduce the rms width of the atomic packet in the initial moment of time, $\rho(0) \equiv \sqrt{\la\rho^2\ra(0)}$, and the de Broglie wavelength $\lambda_{\text{dB}} = 2\pi/\la p\ra$, then 
\bea
\rho(\la z\ra) \approx \frac{\la z\ra}{\rho(0)} \frac{\lambda_{\text{dB}}}{2 \pi} \left(2n + |\ell| + 1\right),
\label{Th}
\eea
For the ground state with $n=\ell=0$, this expression is known as the van Cittert–Zernike theorem \cite{Mandel_Wolf_1995}. Here we have derived it quantum-mechanically by using the exact LG solution to the Schr\"odinger equation. This theorem connects the rms width $\rho(0)$ of the atomic center of mass at a source and $\rho(\la z\ra)$ after the atom propagates over a distance $\la z\ra = \la u\ra_{cm} t$.

\section{Photonic wave packet}\label{PhPack}
For the sake of consistency, we describe the photonic  packets in the same HG$\times$LG basis that is used for atomic wave functions. In fact, one can in principle consider an \textit{arbitrary} photonic wave function in momentum space $\psi^{\gamma}_i(\bk_i,\lambda_i)$, including HG$\times$LG wave function (Eqs. \eqref{HG_mom} and \eqref{LG_mom}), since it is basically a coefficient in the plane-wave expansion of the photonic state \cite{Mandel_Wolf_1995}:
\begin{equation}
\label{superpos}
|\gamma(t)\ra = \sum\limits_{\lambda_i=\pm 1}\int \fr{d^3k}{(2\pi)^3}\,|\bk,\lambda_i\ra\psi^{\gamma}_i(\bk,\lambda_i)e^{-it\omega(\bk)}.
\end{equation}
Thus, we can write the photonic HG$\times$LG packet as
\begin{multline}
\label{HG_LG_phot}
    \psi^{\gamma}(\bk,\lambda_i) = \sqrt{\fr{2\pi\sigma_z^\gamma}{2^{k_\gamma} k_\gamma! \sqrt{\pi}}}\, H_{k_\gamma}\left(\sigma^\gamma_z\left(k_z-\la k_z\ra\right)\right) \exp\Big\{-itk_z - \fr{(\sigma_z^\gamma)^2}{2}\left(k_z-\la k_z\ra\right)^2\Big\} \\
    \times (-1)^{n_\gamma} i^{-\ell} 2 \sqrt{\frac{\pi {n_\gamma}!}{({n_\gamma} + |\ell_\gamma|)!}}  ({\sigma_{\perp}^\gamma})^{\ell_\gamma + 1} k_\perp^{\ell_\gamma} \exp\left\{-\frac{p_\perp^2{\sigma_{\perp}^\gamma}^2}{2}\right\} L_{n_\gamma}^{\ell_\gamma}\left({k_\perp^2 ({\sigma_{\perp}^\gamma})^2}\right)\exp\left\{i(\ell_\gamma+\lambda_i) \phi_k\right\}.
\end{multline}
There are some peculiarities, however, in the definition of the position-dependent photon wave function. Indeed, there is a strongly formulated opinion that the photon wave function in position representation is ill-defined \cite{BLP}, although in quantum optics the following definition of the vectorial photon wave function is accepted rather widely \cite{Mandel_Wolf_1995, fedorov2005spontaneous}:

\begin{equation}
\label{psi_gamma_r}
    \bm \psi^{\gamma}(\br, t) \equiv \fr{1}{\sqrt{\expv{\omega}}}\sum\limits_{\lambda=\pm 1}\int \fr{\dd^3 k}{(2\pi)^3} \bm e_{\lambda_i}(\bk) \sqrt{\omega(\bk)}\psi^{\gamma}_i(\bk, \lambda_i) \:e^{i \bk \cdot \br -i \omega(\bk) t},
\end{equation}
where $\expv{\omega}$ is the mean photon frequency of the superposition \eqref{superpos}. With this definition, $$\expv{\omega}|\bm \psi^{\gamma}(\br, t)|^2$$ represents the energy density of the field. For a quasi-monochromatic packet, the factor $\sqrt{\omega(\bk)}$ can be approximated by $\sqrt{\expv{\omega}}$ and taken out of the integral. Additionally, in the paraxial approximation the polarization vector $e_{\lambda_i}(\bk)$ can be substituted by the transverse polarization vector $\bm \chi_{\lambda_i}$:
\be
\bm \chi_{\pm 1} = \mp \fr{1}{\sqrt{2}}\begin{pmatrix}
    1\\
    \pm i\\
    0
\end{pmatrix}.
\ee

Thus, Eq. \eqref{psi_gamma_r} becomes a standard quantum-mechanical definition of a wave function up to the multiplication by the constant vector:
\begin{equation}
    \bm \psi^{\gamma}(\br, t) \approx \bm \chi_{\lambda_i} \sum\limits_{\lambda_i=\pm 1}\int \fr{\dd^3 k}{(2\pi)^3} \ \psi^{\gamma}_i(\bk, \lambda_i) \:e^{i \bk \cdot \br -i \omega(\bk) t}.
\end{equation}
When the photonic packet is circularly polarized, there is no sum over helicities and it is convenient to define a \textit{scalar} position-space wavefunction as 
$$\bm \psi^{\gamma}(\br, t) \equiv \bm \chi_{\lambda_i} \psi^{\gamma}(\br, t).$$ 
Importantly, this simplification is not suitable for structured optical beams with inhomogeneous polarization induced by spin–orbit coupling. Analyzing the interaction of such photon states with atoms within the present formalism is left for future investigation.

Whereas for quadratic dispersion of a nonrelativistic massive particle one can obtain a closed-form expression for spatial form of HG$\times$LG packet (Eqs. \eqref{HG_pos} and \eqref{LG_pos}), for linear dispersion of the photon $\omega = |\bk|$ it is not so straightforward. Although some cumbersome analytical expressions for symmetric packets can be obtained via the modified Bessel functions \cite{karlovets2018relativistic}, we will not follow this approach here. Instead, we will limit ourselves to the paraxial approximation within which no additional calculations are needed: we can decouple transverse and longitudinal dynamics by expanding the photonic dispersion as
\begin{equation}
    \omega(\bk) = \sqrt{k_\perp^2 + k_z^2} \approx k_z + \fr{k_\perp^2}{2\expv{k_z}}.
\end{equation}
Let us interpret it as a dispersion relation of a nonrelativistic particle with an effective mass $\expv{k_z}$ (and thus with a diffraction time $ \sigma_\perp^2(0) \expv{k_z}/c$) in transverse direction and with an infinite mass (and an infinite diffraction time) in $z$ direction, which completely prevents the packet from longitudinal spreading. The linear dispersion along $z$ axis enables the longitudinal propagation with the velocity $c=1$.

\section{Atom with one electron in external field}\label{AtomHam}

Now take the time-dependent Schr\"odinger equation for an atom with one electron and $Z$ protons in a quantized potential ${\hat{\bm A}}$ obeying the Coulomb gauge with $\hat{A}^0=0, \text{div} {\hat{\bm A}} = 0$,
\bea
& \displaystyle i\partial_t \Psi(\br_e,\br_n,t) = \hat{H}\Psi(\br_e,\br_n,t),\cr
& \displaystyle \hat{H}(\br_e, \br_n, t) = \left(\frac{\left(\hat{\bp}_e - e \hat{{\bm A}}(\br_e, t)\right)^2}{2m} + \frac{\left(\hat{\bp}_n + Z e \hat{{\bm A}}(\br_n, t)\right)^2}{2M} + U(r) + \hat{H}_f\right),
\eea
where $\hat{H}_f = \sum\limits_{\bk,\lambda}\omega_{\bk}\left(\hat{c}_{\bk\,\lambda}^{\dagger}\hat{c}_{\bk\,\lambda} + \frac{1}{2}\right)$. Going to the same variables (\ref{rR}), we get the following Hamiltonian:
\bea
& \displaystyle \hat{H}(\br, \bR, t) = \frac{\hat{\bp}_r^2}{2\mu} + \frac{\hat{\bp}_R^2}{2(m + M)} - e \left(\frac{1}{m}\hat{{\bm A}}(\bR + \br \frac{M}{m+M},t)+\frac{Z}{M}\hat{{\bm A}}(\bR - \br \frac{m}{m+M},t)\right)\cdot \hat{\bp}_r + \cr 
& \displaystyle + \frac{e}{m+M}\left(Z\hat{{\bm A}}(\bR - \br \frac{m}{m+M},t) - \hat{{\bm A}}(\bR + \br \frac{M}{m+M},t)\right)\cdot \hat{\bp}_R + \cr 
& \displaystyle + \frac{e^2}{2m}\hat{{\bm A}}^2(\bR + \br \frac{M}{m+M},t) + \frac{Z^2 e^2}{2M}\hat{{\bm A}}^2(\bR - \br \frac{m}{m+M},t) + U(r) + \hat{H}_f.
\label{Hex}
\eea
One can split the Hamiltonian into the time-independent part $\hat{H}^0$, the time-dependent interaction part, and the relativistic (multiphoton) correction,
\bea
\label{full_Hamiltonian}
& \displaystyle \hat{H} = \hat{H}^0 + \hat{H}^{(\text{int})}+ \hat{H}^{(\text{rel})},\cr
& \displaystyle \hat{H}^0 = \frac{\hat{\bp}_e^2}{2m} + \frac{\hat{\bp}_n^2}{2M} + U(r) + \hat{H}_f = \frac{\hat{\bp}_r^2}{2\mu} + \frac{\hat{\bp}_R^2}{2(m + M)} + U(r) + \hat{H}_f \equiv \hat{H}^{(at)} + \hat{H}_f,\cr
& \displaystyle \hat{H}^{(\text{int})} = -\fr{e}{m}\hat{{\bm A}}(\br_e,t)\cdot \hat{\bp}_e + \fr{Ze}{M}\hat{{\bm A}}(\br_n,t)\cdot \hat{\bp}_n = \cr
& \displaystyle = - e \left(\frac{1}{m}\hat{{\bm A}}\left(\bR + \br \frac{M}{m+M},t\right)+\frac{Z}{M}\hat{{\bm A}}\left(\bR - \br \frac{m}{m+M},t\right)\right)\cdot \hat{\bp}_r + \cr 
& \displaystyle + \frac{e}{m+M}\left(Z\hat{{\bm A}}\left(\bR - \br \frac{m}{m+M},t\right) - \hat{{\bm A}}\left(\bR + \br \frac{M}{m+M},t\right)\right)\cdot \hat{\bp}_R,\cr
& \displaystyle \hat{H}^{(\text{rel})} = \frac{e^2\hat{{\bm A}}^2(\br_e, t)}{2m} + \frac{Z^2e^2\hat{{\bm A}}^2(\br_n, t)}{2M} = \cr
& \displaystyle = \frac{e^2}{2m}\hat{{\bm A}}^2(\bR + \br \frac{M}{m+M},t) + \frac{Z^2 e^2}{2M}\hat{{\bm A}}^2(\bR - \br \frac{m}{m+M},t).
\label{Hspl}
\eea
The Eq.\eqref{full_Hamiltonian} contains the most general form of the interaction Hamiltonian that we use for calculations. We can make several approximations now. 

\textit{Infinitely heavy nucleus.} First, we take the leading non-vanishing contribution in series with the small parameter $m/M \ll 1$ for which
\bea
m+M \approx M, \mu \approx m.
\eea
The result is
\bea
& \displaystyle \hat{H}_{m/M\to 0} = \frac{\hat{\bp}_r^2}{2m} + \frac{\hat{\bp}_R^2}{2M} - e \left(\frac{1}{m}\hat{{\bm A}}(\bR + \br,t)+\frac{Z}{M}\hat{{\bm A}}(\bR,t)\right)\cdot \hat{\bp}_r + \cr 
& \displaystyle + \frac{e}{M}\left(Z\hat{{\bm A}}(\bR,t) - \hat{{\bm A}}(\bR + \br,t)\right)\cdot \hat{\bp}_R + \frac{e^2}{2m}\hat{{\bm A}}^2(\bR + \br,t) + \frac{Z^2 e^2}{2M}\hat{{\bm A}}^2(\bR,t) + U(r) + \hat{H}_f.
\eea

Now we can further make the dipole expansion keeping only the first non-dipole corrections,
\bea
\hat{{\bm A}}(\bR + \br,t) \approx \hat{{\bm A}}(\bR,t) +\left( \br\cdot\nabla_R\right)\, \hat{{\bm A}}(\bR,t),
\eea
and so
\bea
&& \displaystyle \hat{H}_{m/M\to 0} \approx \hat{H}_{m/M\to 0}^{(\text{dip})} + \hat{H}_{m/M\to 0}^{(1)},\cr
&& \displaystyle \hat{H}_{m/M\to 0}^{(\text{dip})} = \frac{\hat{\bp}_r^2}{2m} + \frac{\hat{\bp}_R^2}{2M} - e \left(\frac{1}{m}+\frac{Z}{M}\right)\hat{{\bm A}}(\bR,t)\cdot \hat{\bp}_r + \cr 
&& \displaystyle + \frac{e}{M}\left(Z - 1\right)\hat{{\bm A}}(\bR,t)\cdot \hat{\bp}_R + \frac{e^2}{2}\left(\frac{1}{m} + \frac{Z^2}{M}\right)\hat{{\bm A}}^2(\bR,t) + U(r) + \hat{H}_f,\cr
&& \displaystyle \hat{H}_{m/M\to 0}^{(1)} = -e \br \cdot \nabla_R\, \hat{{\bm A}}(\bR,t) \left(\frac{\hat{\bp}_r}{m} + \frac{\hat{\bp}_R}{M}\right) + \frac{e^2}{2m} \br \cdot \nabla_R\, \hat{{\bm A}}^2(\bR,t).
\eea
Here $\nabla_R = i\hat{\bp}_R$ and within the same accuracy $m/M \to 0$ we can put 
\be
\fr{1}{m} + \fr{Z}{M} \approx \fr{1}{m} + \fr{Z^2}{M} \approx \fr{1}{m}. 
\ee
Note that within the dipole approximation the center of mass couples to the field via the term $\frac{e}{M}\left(Z - 1\right)\hat{{\bm A}}(\bR,t)\cdot \hat{\bp}_R$, non-vanishing only for $Z > 1$.
For the processes with one photon, we have the following interaction part:
\bea
\hat{H}_{m/M\to 0}^{(\text{int})} = \underbrace{- \frac{e}{m}\hat{{\bm A}}(\bR,t)\cdot \hat{\bp}_r + \frac{e}{M}\left(Z - 1\right)\hat{{\bm A}}(\bR,t)\cdot \hat{\bp}_R}_{\text{dipole}} - \underbrace{e \left(\br \cdot \nabla_R\right)\, \hat{{\bm A}}(\bR,t)\cdot \left(\frac{\hat{\bp}_r}{m} + \frac{\hat{\bp}_R}{M}\right)}_{\text{non-dipole}},
\eea
so that the total Hamiltonian can be written as follows:
\bea
&& \displaystyle \hat{H}_{m/M\to 0} \approx \hat{H}_{m/M\to 0}^{(0)} + \hat{H}_{m/M\to 0}^{(\text{int})} + \hat{H}_{m/M\to 0}^{(\text{rel})},\cr
&& \displaystyle \hat{H}_{m/M\to 0}^{(0)} = \frac{\hat{\bp}_r^2}{2m} + \frac{\hat{\bp}_R^2}{2M} + U(r) + \hat{H}_f,\cr
&& \displaystyle \hat{H}_{m/M\to 0}^{(\text{rel})} = \frac{e^2}{2m}\hat{{\bm A}}^2(\bR,t) + \frac{e^2}{2m} \br \cdot \nabla_R\, \hat{{\bm A}}^2(\bR,t).
\eea

{\it Arbitrary} $m/M$. Let us derive the dipole expansion with no assumption on the ratio $m/M$. We start with the exact expression (\ref{Hex}) and get
\bea
& \displaystyle \hat{H} \approx \hat{H}^{(\text{dip})} + \hat{H}^{(1)},\cr
& \displaystyle \hat{H}^{(\text{dip})} = \frac{\hat{\bp}_r^2}{2\mu} + \frac{\hat{\bp}_R^2}{2(m + M)} - e \left(\frac{1}{m}+\frac{Z}{M}\right)\hat{{\bm A}}(\bR,t)\cdot \hat{\bp}_r + \frac{e (Z - 1)}{m+M}\hat{{\bm A}}(\bR,t)\cdot \hat{\bp}_R + \cr 
& \displaystyle + \frac{e^2}{2}\left(\frac{1}{m} + \frac{Z^2}{M}\right)\hat{{\bm A}}^2(\bR,t) + U(r) + \hat{H}_f,
\label{Hdip}
\eea
whereas the non-dipole correction is
\bea
& \displaystyle \hat{H}^{(1)} = \frac{e}{m+M}\left(\frac{Zm}{M} - \frac{M}{m}\right)\left(\br \cdot \nabla_R\right)\, \left(\hat{{\bm A}}(\bR,t)\cdot \hat{\bp}_r\right) - \frac{e}{(m+M)^2}\left(M + mZ\right)\left(\br \cdot \nabla_R\right)\, \left(\hat{{\bm A}}(\bR,t)\cdot \hat{\bp}_R\right) + \cr 
& \displaystyle + \frac{e^2}{2(m+M)}\left(\frac{M}{m}-\frac{mZ^2}{M}\right)\left(\br \cdot \nabla_R\right)\, \hat{{\bm A}}^2(\bR,t)
\eea
The general structure of the interaction part stays the same for arbitrary ratio $m/M$,
\bea
& \displaystyle \hat{H}^{(\text{int})} = \underbrace{- e \left(\frac{1}{m}+\frac{Z}{M}\right)\hat{{\bm A}}(\bR,t)\cdot \hat{\bp}_r + \frac{e (Z - 1)}{m+M}\hat{{\bm A}}(\bR,t)\cdot \hat{\bp}_R}_{\text{dipole}} +\cr 
& \displaystyle + \underbrace{\frac{e}{m+M}\left(\frac{Zm}{M} - \frac{M}{m}\right)\left(\br \cdot \nabla_R\right)\, \left(\hat{{\bm A}}(\bR,t)\cdot \hat{\bp}_r\right) - \frac{e}{(m+M)^2}\left(M + mZ\right)\left(\br \cdot \nabla_R\right)\, \left(\hat{{\bm A}}(\bR,t)\cdot \hat{\bp}_R\right)}_{\text{non-dipole}}.
\eea

\section{General perturbation theory}
\label{Pert_th}
Let us study interaction of an atom with the quantized field in the interaction picture in which both the Hamiltonian and a state depend on time. The state of the whole system,  
\bea
& \displaystyle |\Psi_i\ra = |\psi^{\text{(CM)}}_i, \psi^{e}_i, \gamma_i\ra,
\eea
includes the electron, the center of mass, and the photon field. The evolution from an initial moment of time $t_i$ to some $t_f$ is governed by the solution of the Schr\"odinger equation in the interaction picture
\bea
& \displaystyle |\Psi^{\text{(ev)}}(t_f)\ra = \hat{S}\,|\Psi_i(t_i)\ra,\cr
& \displaystyle \hat{S} = \hat{T}\,\exp\left\{-i\int\limits_{t_i}^{t_f} dt' \hat{H}^{\text{(int)}}(t')\right\},
\label{Sop}
\eea
where $\hat{T}$ is a time ordering operator and
the coordinates $\br$ (relative coordinate) and $\bR$ (CM coordinate) in the interaction Hamiltonian
\bea
\hat{H}^{\text{(int)}}(t') \equiv \hat{H}^{\text{(int)}}(\hat{\br}, \hat{\bR}, t')
\eea
are treated as operators.

The transition matrix element represents the series,
\bea
& \displaystyle S_{fi} = \la\Psi_f(t_f)|\hat{S}\,|\Psi_i(t_i)\ra = \la\Psi_f(t_f)|\Psi_i(t_i)\ra + \la\Psi_f(t_f)|\hat{S}^{(1)}\,|\Psi_i(t_i)\ra + ... = \cr & \displaystyle = \la\Psi_f(t_f)|\Psi_i(t_i)\ra + S_{fi}^{(1)} + S_{fi}^{(2)} + ...,
\eea
where we have taken into account that the in- and out- states are not necessarily orthogonal, $\la\Psi_f(t_f)|\Psi_i(t_i)\ra \ne \delta_{if}$, and the first-order term is
\bea
& \displaystyle S_{fi}^{(1)} = \la\Psi_f(t_f)|\hat{S}^{(1)}\,|\Psi_i(t_i)\ra = -i\int\limits_{t_i}^{t_f} dt' \la\Psi_f(t_f)|\hat{H}^{\text{(int)}}(\hat{\br}, \hat{\bR}, t')|\Psi_i(t_i)\ra
\eea
Now we suppose that the photon field and the atom are {\it not} initially entangled, so that 
\be
|\Psi_i(t_i)\ra = |\psi_i^{(\text{at})}(t_i)\ra \otimes |\gamma_i(t_i)\ra. 
\ee
That is why one can insert the representation of a unity operator on the atomic subspace
\bea
& \displaystyle \hat{1}_{\text{atom}} = \int d^3r d^3 R\, |\br, \bR\ra\la \br, \bR| \label{complat},
\eea
and then we have 
\bea
& \displaystyle \la \br', \bR'|\hat{H}^{\text{(int)}}(\hat{\br}, \hat{\bR}, t') |\br, \bR\ra = \hat{H}^{\text{(int)}}(\br, \bR, t')\,\delta(\br-\br')\delta(\bR-\bR'),
\eea
so we return to the Hamiltonian with the non-operator coordinates. The first-order matrix element now becomes
\bea
& \displaystyle S_{fi}^{(1)} = \la\Psi_f(t_f)|\hat{S}^{(1)}|\Psi_i(t_i)\ra = \cr
& \displaystyle = -i\int dt' d^3r d^3 R\, \la\psi_f^{(\text{at})}(t_f)|\br, \bR\ra \la\gamma_f(t_f)|\hat{H}^{\text{(int)}}(\br, \bR, t')|\gamma_i(t_i)\ra\la \br, \bR|\psi_i^{(\text{at})}(t_i)\ra.
\label{S1X0}
\eea
We denote the atomic wave function in the coordinate representation as 
\be
\psi_i^{(\text{at})}(\br, \bR, t_i) \equiv \la \br, \bR|\psi_i^{(\text{at})}(t_i)\ra.
\ee

The Hamiltonian in the interaction picture is (the coordinate dependence is omitted)
\bea
\label{int_pic_hamilt}
& \displaystyle \hat{H}^{\text{(int)}}(t') = \exp\left\{i t'\hat{H}^0\right\}\hat{H}^{\text{(int)}}(0)\exp\left\{-i t'\hat{H}^0\right\},
\eea
whereas the vectors in the interaction picture are connected with those in the Schr\"odinger picture as
\bea
|\Psi_i(t_i)\ra = \exp\left\{it_i\hat{H}^0\right\}|\Psi_i^{(\text{Sch})}(t_i)\ra = \exp\left\{it_i\hat{H}^{at}\right\}|\psi_i^{(\text{at})}(t_i)\ra^{(\text{Sch})}\exp\left\{it_i\hat{H}_f\right\}|\gamma_i^{(\text{Sch})}(t_i)\ra
\eea
and we have used the representation 
\bea
\hat{H}^0 \equiv \hat{H}^{at} + \hat{H}_f 
\eea
from Eq.(\ref{Hspl}).

In what follows, we omit the signs ``Sch'' and employ only the state vectors in the Schr\"odinger representation, for which
\bea
& \displaystyle S_{fi}^{(1)} = -i\int dt' d^3r d^3 R\, \la\psi_f^{(\text{at})}(t_f)|\br, \bR\ra\,e^{i(t'-t_f)\hat{H}^{(at)}} \cr & \displaystyle
\times \la\gamma_f(t_f)|e^{i(t'-t_f)\hat{H}_f}\hat{H}^{\text{(int)}}(\br, \bR, 0)\,e^{-i(t'-t_i)\hat{H}_f}|\gamma_i(t_i)\ra\,e^{-i(t'-t_i)\hat{H}^{(at)}}\la \br, \bR|\psi_i^{(\text{at})}(t_i)\ra, \cr
& \displaystyle \hat{H}^{(at)} = \frac{\hat{\bp}_r^2}{2\mu} + \frac{\hat{\bp}_R^2}{2(m + M)} + U(r).
\label{S1X2}
\eea

Let the atom be initially free, which implies spatial localization of all the packets involved, so the center-of-mass packet and that of the photon initially do not overlap. We can represent the free state of the CM in the  Schr\"odinger representation as a superposition of plane waves with the momenta $\bP_i$, 
\bea
& \displaystyle |\psi^{(at)}_i(t_i)\ra = \int\fr{d^3P_i}{(2\pi)^3}\,|\bP_i \ra \la\bP_i|\psi^{\text{(CM)}}_i(t_i)\ra|\psi^e_i(t_i)\ra. 
\label{EecmPW}
\eea
where
\bea
\la\bP_i|\psi^{\text{(CM)}}_i(t_i)\ra = \la\bP_i|\psi^{\text{(CM)}}_i\ra\, e^{-i t_i \varepsilon_i^{\text{(CM)}}} \equiv \psi^{\text{(CM)}}_i(\bP_i)\, e^{-i t_i \varepsilon_i^{\text{(CM)}}}
\eea
is the Schr\"odinger wave function in momentum representation with
\bea
\varepsilon_i^{\text{(CM)}}\equiv \varepsilon_i^{\text{(CM)}}(\bP_i) = \frac{\bP_i^2}{2(m+M)}
\eea
being the initial energy of a plane-wave component. As a result
\bea
& \displaystyle \la \br, \bR|\exp\left\{-i\, (t'-t_i)\,\hat{H}^{(at)}\right\}|\psi^{(at)}_i(t_i)\ra = 
\cr & \displaystyle =\la \br|\psi^e_i(0)\ra\, e^{-it'\varepsilon^e_i}\int\fr{d^3P_i}{(2\pi)^3}\, \la\bP_i|\psi^{\text{(CM)}}_i(0)\ra  e^{i\bP_i\cdot\bR-i t'\varepsilon_i^{\text{(CM)}}(\bP_i)} \equiv
\cr & \displaystyle \equiv \psi^e_i(\br)e^{-i t'\varepsilon^e_i} \int\fr{d^3P_i}{(2\pi)^3}\, \, \psi^{\text{(CM)}}_i(\bP_i)\, e^{i\bP_i\cdot\bR-i t'\varepsilon_i^{\text{(CM)}}(\bP_i)} = \psi^e_i(\br)e^{-i t'\varepsilon^e_i}\psi^{\text{(CM)}}_i(\bR,t'),
\label{EatPW}
\eea

where we have used
\bea
\la\bR|\bP_i\ra = e^{i\bP_i\cdot\bR}.
\eea
Note that the dependence of the phases on $t_i$ has been canceled.
Here
\bea
\psi^{\text{(CM)}}_i(\bR,t') = \int\fr{d^3P_i}{(2\pi)^3}\, \psi^{\text{(CM)}}_i(\bP_i)  e^{i\bP_i\cdot\bR-i t'\varepsilon_i^{\text{(CM)}}(\bP_i)} = \psi_{||}^{(k)}(Z, t')\psi_{\perp}^{(n,\ell)}({\bm \rho}, t')
\eea
is the HG$\times$LG wave function of the atomic CM from Eq.(\ref{HG_LG}) with ${\bm R}=\{{\bm \rho}, Z\}$.

If there is a photon packet only in the initial state,
\bea
|\gamma_i(t_i)\ra = \sum\limits_{\lambda_i=\pm 1}\int \fr{d^3k_i}{(2\pi)^3}\,|\bk_i,\lambda_i\ra\la \bk_i,\lambda_i|\gamma_i\ra\,e^{-it_i\omega_i},\ |\bk_i,\lambda_i\ra = \hat{c}_{\bk_i,\lambda_i}|0\ra,
\eea
 a photon absorption occurs and so the final photon state is vacuum, $|\gamma_f(t_f)\ra = |0\ra$. Then
\bea
& \displaystyle \la\gamma_f(t_f)|e^{i(t'-t_f)\hat{H}_f}\hat{H}^{\text{(int)}}(\br, \bR, 0)\,e^{-i(t'-t_i)\hat{H}_f}|\gamma_i(t_i)\ra = \cr & \displaystyle = \sum\limits_{\lambda_i}\int\fr{d^3k_i}{(2\pi)^3}\,\psi^{\gamma}_i(\bk_i,\lambda_i)\,\left(-\fr{e}{m}\,{\bm A}_{\bk_i,\lambda_i}(\br_e)\cdot\hat{\bp}_e + \fr{Ze}{M}\,{\bm A}_{\bk_i,\lambda_i}(\br_n)\cdot\hat{\bp}_n\right)\,e^{-it'\omega_i(\bk_i)},
\eea
where 
$\psi^{\gamma}_i(\bk_i,\lambda_i) \equiv \la\bk_i,\lambda_i|\gamma_i\ra$ is a scalar part of the initial photon wave function and
we have used that
\bea
& \displaystyle \la 0|\hat{H}^{(\text{int})}(\br,\bR,0)|\bk_i,\lambda_i\ra = -\fr{e}{m}\,{\bm A}_{\bk_i,\lambda_i}(\br_e)\cdot\hat{\bp}_e + \fr{Ze}{M}\,{\bm A}_{\bk_i,\lambda_i}(\br_n)\cdot\hat{\bp}_n,\cr
& \displaystyle {\bm A}_{\bk,\lambda}(\br) = \frac{1}{\sqrt{2\omega_{\bk}}}\,{\bm e}_{\lambda}(\bk)\, e^{i\br\cdot\bk}.
\eea

When photons exist both in the initial and in the final states, we find (see Eq.(\ref{S1X2}))
\bea
& \displaystyle \la\gamma_f(t_f)|e^{i(t'-t_f)\hat{H}_f}\hat{H}^{\text{(int)}}(\br, \bR, 0)\,e^{-i(t'-t_i)\hat{H}_f}|\gamma_i(t_i)\ra = \cr & \displaystyle = \sum\limits_{\lambda_i,\lambda_f}\int\fr{d^3k_i}{(2\pi)^3}\fr{d^3k_f}{(2\pi)^3}\,\psi^{\gamma}_i(\bk_f,\lambda_f)^* \,e^{it'\omega_f(\bk_f)}\la \bk_f,\lambda_f|\hat{H}^{(\text{int})}(\br,\bR,0)|\bk_i,\lambda_i\ra\psi^{\gamma}_i(\bk_i,\lambda_i) \,e^{-it'\omega_i(\bk_i)}.
\eea
Clearly
$\la \bk_f,\lambda_f|\hat{H}^{(\text{int})}(\br,\bR,0)|\bk_i,\lambda_i\ra = 0,
$
whereas the quadratic in ${\bm A}$ part of the Hamiltonian is not vanishing,
\bea
& \displaystyle \la \bk_f,\lambda_f|\hat{H}^{(\text{rel})}(\br,\bR,0)|\bk_i,\lambda_i\ra \propto \la \bk_f,\lambda_f|\hat{{\bm A}}^2(\br,t)|\bk_i,\lambda_i\ra = \cr 
& \displaystyle =(2\pi)^3\delta_{\lambda_i\lambda_f}\delta(\bk_i-\bk_f)\sum\limits_{\lambda}\int\fr{d^3k}{(2\pi)^3} \left(\left|{\bm A}_{\bk\lambda}(\br,t)\right|^2\equiv\fr{1}{2\omega_{\bk}}\right) + 2{\bm A}_{\bk_i\lambda_i}(\br,t)\cdot{\bm A}^*_{\bk_f\lambda_f}(\br,t).
\eea
The first term here is divergent and should be omitted (it is the infinite energy of vacuum fluctuations), while the second term enables scattering in the first order of perturbation theory discussed below and in Sec.~2C in the main text.

Referring again to absorption, when both the atom and the absorbed photon are described as wave packets we arrive at
\bea
& \displaystyle S_{fi}^{(1)} = -i\sum\limits_{\lambda_i=\pm 1}\int dt' d^3r d^3 R\,\fr{d^3P_i}{(2\pi)^3}\,\fr{d^3P_f}{(2\pi)^3}\,\fr{d^3k_i}{\sqrt{2\omega_{\bk_i}}(2\pi)^3}\,\exp\left\{-it'\left(\varepsilon^{\text{(CM)}}_i(\bP_i)+\varepsilon^{e}_i+\omega_i-\varepsilon^{\text{(CM)}}_f(\bP_f)-\varepsilon^e_f\right)\right\}\cr
& \displaystyle
\left(\psi_{f}^e(\br)\right)^*\,\left(\psi_{f}^{\text{(CM)}}({\bP}_f)\right)^*\,\exp\left\{-i\bR\cdot\bP_f\right\}\, {\bm e}_{\lambda_i}(\bk_i)\cdot\Bigg(-\fr{e}{m}\,\exp\left\{i\bk_i\cdot\br\fr{M}{m+M}\right\}\left(\hat{\bp}_r + \fr{m}{m+M}\bP_i\right) + \cr & \displaystyle + \frac{Ze}{M}\,\exp\left\{-i\bk_i\cdot\br\fr{m}{m+M}\right\}\left(-\hat{\bp}_r + \fr{M}{m+M}\bP_i\right)\Bigg)\, \psi^e_i(\br)\psi^{\text{(CM)}}_i({\bP}_i)\psi^{\gamma}_i({\bk}_i,\lambda_i)\, \exp\left\{i\bR\cdot(\bP_i+\bk_i)\right\}.
\label{S1X1}
\eea
As all the dependence on $t_i, t_f$ {\it has been canceled}, we can safely put $t_i\to-\infty, t_f\to \infty$ and get the conservation laws for each plane-wave component
\bea
& \displaystyle \int\limits_{-\infty}^{\infty} dt'\,\exp\left\{-it'\left(\varepsilon^{\text{(CM)}}_i(\bP_i)+\varepsilon^{e}_i+\omega_i-\varepsilon^{\text{(CM)}}_f(\bP_f)-\varepsilon^e_f\right)\right\} = 2\pi \delta \left(\varepsilon^{\text{(CM)}}_i(\bP_i)+\varepsilon^{e}_i+\omega_i-\varepsilon^{\text{(CM)}}_f(\bP_f)-\varepsilon^e_f\right),\cr
& \displaystyle \int d^3R\,\exp\left\{i\bR\cdot(\bP_i+\bk_i- \bP_f)\right\} = (2\pi)^3\delta(\bP_i+\bk_i- \bP_f).
\eea
Note that the energy conservation law includes the energies of the electron, of the CM, and of the photon packet, whereas the momentum conservation law only includes momenta of the CM and of the photon. This makes photon absorption without the change of the electronic state kinematically forbidden.

Let us now consider the second-order terms in $S_{fi}$, which produce scattering. We will only discuss transitions between the plane-wave states of the CM and the photon because the generalization to wave packets is done similarly to the above procedure. The electronic state can, in principle, also be changed by scattering: we call such a process \textit{inelastic scattering}, although technically even scattering without a change of the electronic state is inelastic due to atomic recoil. 

A general second-order S-matrix term reads
\begin{equation}
    \hat{S}^{(2)} = \left(-i\right)^2\fr{1}{2}\hat{T}\left[\int_{t_i}^{t_f} \hat{H}^{\text{(int)}}(t') \dd t'\right]^2 = \left(-i\right)^2 \int_{t_i}^{t_f} \dd t' \hat{H}^{\text{(int)}}(t') \int_{t_i}^{t'} \dd t'' \hat{H}^{\text{(int)}}(t'').
\end{equation}

Consider a transition from some initial state $\ket{i}$ to $\ket{f}$ through a set of intermediate states $\ket{y}$ of the whole system (electron, CM, and field), which are eigenstates of $\hat{H}^0$. Inserting the unity operator $\hat 1 = \sum_y\ket{y}\bra{y}$ and using Eq. \eqref{int_pic_hamilt}, we get
\begin{multline}
    S_{fi}^{(2)} = \sum_y\left(-i\right)^2 \int_{t_i\to -\infty}^{t_f\to \infty} \dd t' e^{i(\omega_f-\omega_y)t'} \bra{f}\hat{H}^{\text{(int)}}(0)\ket{y} \int_{t_i\to -\infty}^{t'} \dd t''e^{i(\omega_y-\omega_i)t''} \bra{y}\hat{H}^{\text{(int)}}(0)\ket{i} = \\
   =\sum_y \left(-i\right)^22\pi i\delta(\omega_f-\omega_i) \fr{1}{\omega_i-\omega_y} \bra{f}\hat{H}^{\text{(int)}}(0)\ket{y} \bra{y}\hat{H}^{\text{(int)}}(0)\ket{i}.
\end{multline}
Thus, the matrix element becomes
\begin{multline}
    \label{S(2)}
     S_{fi}^{(2, PW)} \equiv \bra{\bP_f, \bk_f, \psi_f^e}\hat{S}^{(2)}\ket{\bP_i, \bk_i, \psi_i^e} = - 2\pi i\delta\left(\varepsilon^{\text{(CM)}}_i(\bP_i)+\varepsilon^{e}_i+\omega_i-\varepsilon^{\text{(CM)}}_f(\bP_f) - \varepsilon^e_f(\bp') - \omega_f\right) \\
    \times \sum_{\tilde{\psi}^e} \int \fr{\dd^3 \tilde P}{(2\pi)^3} \Biggl[ \fr{\bra{\bP_f, \bk_f, \psi_f^e}\hat{H}^{\text{(int)}}(0)\ket{\tilde{\bP}, \tilde{\psi}^e} \bra{\tilde{\bP}, \tilde{\psi}^e}\hat{H}^{\text{(int)}}(0)\ket{\bP_i, \bk_i, \psi_i^e}}{\varepsilon_i^e + \omega_i + \fr{\bP_i^2}{2(m+M)} - \fr{\tilde{\bP}^2}{2(m+M)} - \tilde{\varepsilon}^e} \\
    + \fr{\bra{\bP_f, \bk_f, \psi_f^e}\hat{H}^{\text{(int)}}(0)\ket{\tilde{\bP}, \bk_i, \bk_f, \tilde{\psi}^e} \bra{\tilde{\bP}, \bk_i, \bk_f, \tilde{\psi}^e}\hat{H}^{\text{(int)}}(0)\ket{\bP_i, \bk_i, \psi_i^e}}{\varepsilon_i^e +\fr{\bP_i^2}{2(m+M)} - \fr{\tilde{\bP}^2}{2(m+M)} - \omega_f - \tilde{\varepsilon}^e}. \Biggr]
\end{multline}
Here, $\bk_{i, f}$ is a multi-index containing both the photon momentum and the helicity $\lambda_{i, f}$. The only two possible types of intermediate states are the 2-particle states of the CM and the electron $\ket{\tilde{\bP}, \tilde{\psi}^e}$ and the states with two additional photons $\ket{\tilde{\bP}, \bk_i, \bk_f, \tilde{\psi}^e}$. The unity operator on the atomic subspace is here expanded over discrete electronic states and continuous spectrum of the CM as
\bea
    \hat 1_{\text{atom}} = \sum_{\tilde{\psi}^e} \int \dd^3 \tilde P /(2\pi)^3\ket{\tilde{\bP}, \tilde{\psi}^e} \bra{\tilde{\bP}, \tilde{\psi}^e}
\eea
and $\tilde \varepsilon^e$ is an energy of an intermediate electronic state. Now, let us express the matrix elements of the interaction Hamiltonian in terms of the transition currents~$(2)$:
\begin{equation}
    \bra{\tilde{\bP}, \tilde{\psi}^e}\hat{H}^{\text{(int)}}(0)\ket{\bP_i, \bk_i, \psi_i^e}
   = \fr{e}{\sqrt{2\omega_{i}}}(2\pi)^3 \delta(\bP_i + \bk_i-\tilde{\bP})\: \bm e_{\lambda_i}(\bk_i) \cdot \bm{J}_{i \tilde{\psi}^e}(\bk_i, \bP_i);
\end{equation}

\begin{multline}
    \bra{\bP_f, \bk_f, \psi_f^e}\hat{H}^{\text{(int)}}(0)\ket{\tilde{\bP}, \tilde{\psi}^e}  =  \bra{\tilde{\bP}, \tilde{\psi}^e}\hat{H}^{\text{(int)}}(0)\ket{\bP_f, \bk_f, \psi_f^e}^* \\
    =\fr{e}{\sqrt{2\omega_{f}}}(2\pi)^3 \delta(\bP_f + \bk_f-\tilde{\bP})\:\bm e_{\lambda_f}^*(\bk_f) \cdot {\bm{J}^*_{f \tilde{\psi}^e}}(\bk_f, \bP_f);
\end{multline}

\begin{equation}
    \bra{\tilde{\bP}, \bk_i, \bk_f, \tilde{\psi}^e}\hat{H}^{\text{(int)}}(0)\ket{\bP_i, \bk_i, \psi_i^e}
   = \fr{e}{\sqrt{2\omega_{f}}}(2\pi)^3 \delta(\bP_i -\bk_f -\tilde{\bP})\:\bm e_{\lambda_f}^*(\bk_f) \cdot {\bm{J}^*_{\tilde{\psi}^e i}}(\bk_f, \tilde{\bP});
\end{equation}

\begin{equation}
    \bra{\bP_f, \bk_f, \psi_f^e}\hat{H}^{\text{(int)}}(0)\ket{\tilde{\bP}, \bk_i, \bk_f, \tilde{\psi}^e}
   = \fr{e}{\sqrt{2\omega_{i}}}(2\pi)^3 \delta(\tilde{\bP} +\bk_i -\bP_f)\:\bm e_{\lambda_i}(\bk_i) \cdot {\bm{J}_{\tilde{\psi}^e f}}(\bk_i, \tilde{\bP}).
\end{equation}

{\red Thus, the  plane-wave amplitude reads
\begin{multline}
\label{S2_spec}
     S_{fi}^{(2, PW)} = -i(2\pi)^4 \delta \left(\varepsilon^{\text{(CM)}}_i(\bP_i)+\varepsilon^{e}_i+\omega_i-\varepsilon^{\text{(CM)}}_f(\bP_f)-\omega_f -\varepsilon^e_f\right)
    \delta(\bP_i+\bk_i-\bP_f-\bk_f)  \fr{e^2}{\sqrt{4\omega_{i}\omega_{f}}} \\
    \times \sum_{\tilde{\psi}^e}\Biggl[ \fr{\left(\bm e_{\lambda_i}(\bk_i) \cdot \bm{J}_{i \tilde{\psi}^e }(\bk_i, \bP_i)\right) \left(\bm e_{\lambda_f}(\bk_f) \cdot {\bm{J}_{f \tilde{\psi}^e}}(\bk_f, \bP_f)\right)^*}{\varepsilon_i^e-\tilde{\varepsilon}^e+\omega_i+\fr{\bP_i^2}{2(m+M)} - \fr{(\bP_i+\bk_i)^2}{2(m+M)}} + \fr{\left(\bm e_{\lambda_i}(\bk_i) \cdot \bm{J}_{\tilde{\psi}^e f}(\bk_i, \bP_f)\right) \left(\bm e_{\lambda_f}(\bk_f) \cdot {\bm{J}_{\tilde{\psi}^e i}}(\bk_f, -\bP_i)\right)^*}{\varepsilon_i^e-\tilde{\varepsilon}^e+\fr{\bP_i^2}{2(m+M)} - \fr{(\bP_i-\bk_f)^2}{2(m+M)}-\omega_f}\Biggr].
\end{multline}
Here $\tilde{\psi}^e$ denotes the intermediate electronic states. As before, the momentum conservation includes only the CM momenta $\bP_{i, f}$ and the photon momenta $\bk_{i, f}$, whereas the energy conservation law includes the electron energies $\varepsilon^e_{i, f}$, the CM energies $\varepsilon^{\text{(CM)}}_{i, f} = \bP_{i_f}^2 / 2(m+M)$ and the photon frequencies $\omega_{i, f}=|\bk_{i, f}|$ . The poles of the amplitude  occur when the photon energy matches the electronic transition energy with the recoil shift taken into account (the first term in \eqref{S2_spec}). The second term produces the pole lying in the unphysical domain.

There is, however, another contribution to the scattering, arising in the first order of perturbation theory from the Hamiltonian
\begin{equation}
     \hat{H}^{(\text{rel})} = \frac{e^2}{2m}\hat{{\bm A}}^2(\bR + \br \frac{M}{m+M},t) + \frac{Z^2 e^2}{2M}\hat{{\bm A}}^2(\bR - \br \frac{m}{m+M},t)
\end{equation}
The corresponding transition amplitude is
\begin{multline}
\label{S15}
    S_{fi}^{(1.5, PW)} \equiv\bra{\bP_f, \bk_f, \psi_f^e}\hat{S}^{(1.5)}\ket{\bP_i, \bk_i, \psi_i^e} \\
    =-i(2\pi)^4 \delta \left(\varepsilon^{\text{(CM)}}_i(\bP_i)+\varepsilon^{e}_i+\omega_i-\varepsilon^{\text{(CM)}}_f(\bP_f)-\omega_f -\varepsilon^e_f\right)\delta(\bP_i+\bk_i-\bP_f-\bk_f) \\
    \times \fr{ \left(\bm e_{\lambda_i}(\bk_i) \cdot \bm e_{\lambda_f}^*(\bk_f)\right)}{\sqrt{\omega_{i}\omega_{f}}}\fr{e^2}{2m}\Bigg[\int (\psi_f^e(\br))^* e^{i(\bk_i-\bk_f)\cdot \br \fr{M}{m+M}} \psi_i^e(\br) \dd^3 r 
    +\fr{Z^2m}{M} (\psi_f^e(\br))^* e^{-i(\bk_i-\bk_f)\cdot \br \fr{m}{m+M}} \psi_i^e(\br) \dd^3 r \Bigg].
\end{multline}
Interestingly, this amplitude also allows for inelastic scattering, but {\it only beyond the dipole approximation}, otherwise the integral in Eq. \eqref{S15} vanishes for $\psi_i^f(\br) \neq \psi_i^e(\br)$.}

\section{Electron transitions}\label{ElTr}
\begin{table}[h!]
\centering
\renewcommand{\arraystretch}{1.5}
\begin{tabular}{|c|c|c|}
\hline
\textbf{Final state} & \textbf{Transition current} $j^{(\nu=1,3)}_{\lambda_i}$ & \textbf{Transition current} $j^{(\nu=2,4)}_{\lambda_i} / (P_f)_{\bk\lambda}$ \\
\hline
$1s$ & $0$ & $\fr{16}{\left(4 + (a_0k\beta)^2\right)^2}$ \\
\hline
$2s$ & $0$ & $\fr{256\sqrt{2} (a_0 k \beta)^2}{\left(9 + 4(a_0k\beta)^2\right)^3}$  \\
\hline
$2p$ ($m_f=-1$) & $-\fr{16i(\lambda-1)}{a_0\left(9+4(a_0k \beta)^2\right)^2}$ & $0$ \\
$2p$ ($m_f=0$)  & $0$   & $\fr{384i\sqrt{2} (a_0 k \beta)}{\left(9 + 4(a_0k\beta)^2\right)^3}$ \\
$2p$ ($m_f=+1$) & $-\fr{16i(\lambda+1)}{a_0\left(9+4(a_0k \beta)^2\right)^2}$ & $0$ \\
\hline
$3s$  & $0$ & $\fr{432\sqrt{2}(a_0k\beta)^2\left(16+27(a_0k\beta\right)^2)}{\left(16+9(a_0k\beta)^2\right)^4}$ \\
\hline
$3p$ ($m_f=-1$) & $-\fr{48i(8+9(a_0 k \beta)^2)(\lambda-1)}{a_0\left(16+9(a_0k\beta)^2\right)^3}$ & $0$ \\
$3p$ ($m_f=0$)  & $0$   & $\fr{864i\sqrt{2}a_0k\beta(16+27(a_0k\beta)^2)}{(16+9(a_0k\beta)^2)^4}$ \\
$3p$ ($m_f=+1$) & $-\fr{48i(8+9(a_0 k \beta)^2)(\lambda+1)}{a_0\left(16+9(a_0k\beta)^2\right)^3}$ & $0$ \\
\hline
$3d$ ($m_f=-2$) & $0$ & $0$ \\
$3d$ ($m_f=-1$) & $\fr{288k\beta(\lambda-1)}{\left(16+9(a_0k\beta)^2\right)^2}$ & $0$ \\
$3d$ ($m_f=0$)  & $0$   & $-\fr{6912\sqrt{6}(a_0k\beta)^2}{\left(16+9(a_0k\beta)^2\right)^4}$ \\
$3d$ ($m_f=+1$) & $\fr{288k\beta(\lambda+1)}{\left(16+9(a_0k\beta)^2\right)^2}$ & $0$ \\
$3d$ ($m_f=+2$) & $0$ & $0$ \\
\hline
\end{tabular}
\caption{Transition currents $j_{\lambda_i}$ from $1s$ to $n\ell$ states with different $m$ projections. $\beta = \fr{M}{m+M}$ for $\nu=1, 2$ and $\beta = -\fr{m}{m+M}$ for $\nu=3, 4$.}
\label{tab:matrix_elements}
\end{table}

Generally, the non-relativistic wave function for a hydrogen-like atom is given by:
\be
\psi_{nlm}(r, \theta, \phi) = R_{nl}(r) Y_l^m(\theta, \phi),
\ee
where
\be
Y_l^m(\theta, \phi) = \sqrt{\frac{(2l+1)}{4\pi} \cdot \frac{(l-m)!}{(l+m)!}} P_l^m(\cos\theta) e^{im\phi}
\ee
is the spherical harmonic, and $R_{nl}(r)$ is the radial wave function. The radial wave function $R_{nl}(r)$ for the hydrogen-like atom is \cite{messiah2014quantum}
\be
R_{nl}(r) = \sqrt{\left(\frac{2}{na_0}\right)^3 \frac{(n-l-1)!}{2n(n+l)!}} e^{-\frac{r}{na_0}} \left(\frac{2r}{na_0}\right)^l L_{n-l-1}^{2l+1}\left(\frac{2r}{na_0}\right),
\ee
where $a_0=\fr{4\pi}{Ze^2\mu}$ is the effective Bohr radius including the effect of reduced mass and the nucleus charge $Z$.

The electron wave functions are transformed to a coordinate system with quantization axis being aligned with $\bk$ as follows \cite{scholz2014absorption, Varshalovich}: 
\begin{equation}
\label{rotated_psi}
    \psi_{n\ell m}(\bm r) = \sum_{m'} D^{\ell*}_{m m'}(\phi_k, \theta_k, 0) \psi_{n\ell m'}(\bm r'),
\end{equation}
where $D^{\ell}_{m m'}(\phi_k, \theta_k, 0) = e^{-im\phi_k} d^\ell_{mm'}(\theta_k)$. These matrix elements are multiplied by the polarization vector
\begin{equation}
    \bm e_{\lambda_i}(\bk) = \sum_{\sigma=0, \pm1} e^{-i\sigma \phi_k} d^1_{\sigma \lambda}(\theta_k)\bm \chi_\sigma.
\end{equation}
It is convenient to put the polarization vector under the integral and use the fact that the dot products are invariant under rotations. When the wave functions are rotated to the quantization axis along $\bk$, the nabla operator is also transformed to the new coordinates, and the polarization vector $\bm e_{\lambda_i}(\bk)$ transforms as \cite{scholz2014absorption, EPJC2023, EPJC2024}
\begin{equation}
    \bm e_{\lambda_i}(\bk) \to \bm \chi_{\lambda_i}.
\end{equation}
The dot product becomes
\begin{align}
    \bm e_{\lambda_i}(\bk)\cdot \hat{\bp}_r &\rightarrow  (\hat{p}_r)_{\lambda},
\end{align}
where $\lambda=\pm 1$ now refers to the \textit{covariant} spherical components of the vector
\begin{gather}
    (\hat{p}_r)_{\lambda_i} 
    = \fr{i\lambda}{\sqrt{2}}e^{i\lambda \phi} \left(\sin \theta \hat \partial_r + \fr{\cos \theta}{r}\hat\partial_\theta + \fr{i\lambda}{r \sin \theta}\hat\partial_\phi\right).
\end{gather}
There is one more dot product involving azimuthal angles in the expression for the evolved state:

\begin{multline}
     \bm{e}_{\lambda}(\bk)\cdot\bP_i  = \bm{e}_{\lambda}(\bk)\cdot(\bP_f - \bk)= \bm{e}_{\lambda}(\bk)\cdot\bP_f 
    \\
    =\frac{P_{f\perp}}{\sqrt{2}}e^{i(\phi_k-\phi_f)}d_{-1\lambda}^1(\theta_k)+ P_{fz} d_{0\lambda}^1(\theta_k) -\frac{P_{f\perp}}{\sqrt{2}}e^{-i(\phi_k-\phi_f)}d_{1\lambda}^1(\theta_k) \equiv (P_f)_{\bk\lambda}
\end{multline}
(see \cite{Pavlov2024}). Further it will be shown that the phases $(\phi_k-\phi_i)$ will be reduced due to delta functions. 

Thus, the electronic transition integrals  can be written as
\begin{equation}
\label{ej}
    \bm e_{\lambda_i}(\bk)\cdot\bm j^{(1, 2, 3, 4)}(\bk) = e^{-i(m_f-m_i)\phi_k}\sum_{m_i', m_f'}  j^{(1, 2, 3, 4)}_{\lambda} d^{\ell_i}_{m_im_i'}(\theta_k) d^{\ell_f}_{m_fm_f'}(\theta_k),
\end{equation}
where
\begin{multline}
\label{j_lambda}
j^{(1,2,3,4)}_{\lambda_i}=\\
=\int d^3r\,(\psi_{f}^e(\br))^* \Big\{e^{ikz\fr{M}{m+M}}(\hat{p}_r)_{\lambda_i}, e^{ikz\fr{M}{m+M}}(P_f)_{\bk\lambda},e^{-ikz\fr{m}{m+M}}(\hat{p}_r)_{\lambda_i}, e^{-ikz\fr{m}{m+M}}(P_f)_{\bk\lambda} \Big\}\psi^e_i(\br)
\end{multline}
(see \cite{scholz2014absorption, baturin2024conversion}).

The azimuthal dependence $\bm j^{(1, 2,3, 4)}(\bk)\cdot  \bm e_{\lambda_i}(\bk) \propto e^{-i(m_f-m_i) \phi_k}$ is a very important property since it enables the OAM conservation.

\section{Delta function in cylindrical coordinates}\label{deltacyl}

One can separate the momentum delta function into transverse and longitudinal parts as
\begin{equation}
    \delta^{(3)}(\bP_i+\bk- \bP_f) =  \delta^{(2)}(\bP_{i\perp}+\bk_{\perp}- \bP_{f\perp})\delta(P_{iz}+k_{z}-P_{fz}).
\end{equation}
The transverse part can be then rewritten via azimuthal angles  in cylindrical coordinates
\begin{multline}
\label{delta_perp}
\delta^{(2)}(\bP_{i\perp}+\bk_{i\perp}- \bP_{f\perp}) =\\
\frac{\Theta(P_{i\perp},k_{i\perp},P_{f\perp})}{2\Delta} \Big(\delta\left(\phi_i - \phi_f + \alpha\right)\delta\left(\phi_k - \phi_f - \gamma\right) + \delta\left(\phi_i - \phi_f - \alpha\right)\delta\left(\phi_k - \phi_f + \gamma\right)\Big) =\\
= \frac{\Theta(P_{i\perp},k_{i\perp},P_{f\perp})}{2\Delta} \Big(\delta\left(\phi_i - \phi_k - (\beta-\pi)\right)\delta\left(\phi_k - \phi_f - \gamma\right) + \delta\left(\phi_i - \phi_k + (\beta-\pi)\right)\delta\left(\phi_k - \phi_f + \gamma\right)\Big),
\end{multline}
where $\Delta$ is an area of the triangle with the legs $P_{i\perp}, P_{f\perp}, k_{\perp}$ and the angles $\alpha, \beta, \gamma\, (\alpha + \beta + \gamma = \pi)$,
\begin{gather}
\Delta = \frac{1}{2} P_{f\perp}P_{i\perp}\sin\alpha = \frac{1}{2} k_{\perp}P_{i\perp}\sin\beta = \frac{1}{2} P_{f\perp}k_{\perp}\sin\gamma \\
\alpha = \arccos\left\{\frac{P_{f\perp}^2 + P_{i\perp}^2 - k_{\perp}^2}{2P_{f\perp}P_{i\perp}}\right\},\, \beta = \arccos\left\{\frac{P_{i\perp}^2 + k_{\perp}^2 - P_{f\perp}^2}{2k_{\perp}P_{i\perp}}\right\},\, \gamma = \arccos\left\{\frac{P_{f\perp}^2 + k_{\perp}^2 - P_{i\perp}^2}{2k_{\perp}P_{f\perp}}\right\}.
\end{gather}

The function $\Theta(p_{\perp}, k_{\perp}, p'_{\perp})$ equals 1 when the legs satisfy the triangle rules, 
\begin{equation}
    P_{f\perp} < P_{i\perp} + k_\perp, \; P_{i\perp} < P_{f\perp} + k_\perp, \;  k_\perp < P_{i\perp} + P_{f\perp},
\end{equation}
and vanishes otherwise. More details on this expansion can be found, for instance, in Ref. \cite{ivanov2011colliding} or in Supplementary of Ref. \cite{karlovets2025attosecond}.

{\red Such an expansion is useful for the explicit verification of OAM conservation. Indeed, the evolved state wave function~$(5)$ in the main text with the aid of Eq.~\eqref{delta_perp} can be written as 
\begin{multline}
\label{Transormed_Ev_state_CM}
\psi^{\text{(CM, ev)}}(\bP_f) =-i e^{i (\ell_\gamma+\lambda_i+ \ell^{(\text{CM})}_i+m_i-m_f)\phi_f}e^{-i t_f \varepsilon^{\text{(CM)}}_f}e\sum_{\lambda_i} \fr{M+m}{(2\pi)^2}\int  \Theta(P_{i\perp},k_{i\perp},P_{f\perp}) \fr{k_\perp }{\sqrt{2\omega_k}2\Delta} d k_{i\perp} d k_{iz} \mathcal J_{\lambda_i}(\bk_i, \bP_f)\\
\times \tilde\psi^{\text{(CM)}}(P_{\perp i}, P_{iz})\tilde\psi^{\gamma}(k_\perp, k_z, \lambda_i) e^{i(\ell_\gamma+\lambda_i+m_i-m_f)\gamma - i \ell^{(\text{CM})}_i\alpha-ik_\perp b\cos(\phi_b-\phi_f-\gamma)} +\{\text{the same with }\alpha\to-\alpha,\,\gamma\to-\gamma\},
\end{multline}

where $P_{iz} = P_{fz}-k_z$ is implied and $P_{i\perp}=\sqrt{P_{\perp f}^2-k_z^2+2P_{fz}k_z+2(M+m)(\varepsilon^e_f-\omega-\varepsilon_i^e)}$.
Clearly, for a head-on collision with {\it the vanishing impact parameter} $\bm b\to 0$ there is no dependence on the angle $\phi_f$ in the integrand, thus the evolved state becomes an eigenfunction of the OAM z-projection operator. Although Eq.\eqref{Transormed_Ev_state_CM} explicitly shows the OAM conservation, its numerical performance is hampered. Sec. X of this Supplementary Material introduces an alternative parametrization of Eq~$(5)$ that is more stable for numerical work.}

\section{OAM distribution in the evolved state of CM: analytics}
{\red According to Sec II B of the main text, after a collision with non-vanishing impact parameter $b$ the evolved state of CM can be estimated as
\begin{equation}
\label{estimated_dist}
\ket{\psi^{\text{(CM, ev)}}} \propto 
\sum \limits_{m=-\infty}^{\infty}(-i)^m J_m(b / \sigma_\perp^\gamma) 
\times I_{\ell_0+m} \left(\sigma^{(\text{CM})}_\perp / \sigma_\perp^\gamma \right) \ket{\ell_0+m},
\end{equation}
where $\ell_0$ is the OAM transferred at $b=0$. For most relevant regime in experiment, where $b \ll \sigma_\perp^\gamma$ and $\sigma^{(\text{CM})}_\perp \ll \sigma_\perp^\gamma$, the asymptotics of Bessel functions can be used. In this case, {\it the probability for the CM to carry the OAM value} of $\ell_0+m$ is approximately
\begin{equation}
    P(\ell_0+m) = \Biggl[  \fr{\left(\fr{b}{2\sigma_\perp^\gamma}\right)^{|m|}}{|m|!}  \fr{\left(\fr{\sigma_\perp^{(\text{CM})}}{2\sigma_\perp^\gamma}\right)^{|\ell_0+m|}}{|\ell_0+m|!}  \Biggr]^2.
\end{equation}
From this, we obtain the ratio
\begin{equation}
    P(\ell_0-1) / P(\ell_0) = (\ell_0\: b/\sigma_\perp^{(\text{CM})})^2
\end{equation}
and the similar for higher orders, continuing until $\ell_0-m=0$. Thus, if $b<\sigma_\perp^{(\text{CM})}$, the OAM values below $\ell_0$ - but still positive - are suppressed. Conversely, if $b>\sigma_\perp^{(\text{CM})}$, the OAM values below $\ell_0$ become more probable than the OAM transferred for head-on absorption.

However, this trend reverses once $\ell_0-m<0$, as the distribution decays sharply:
\begin{equation}
    P(-1) / P(0) = \left(\fr{b}{2\sigma_\perp^\gamma}\fr{\sigma_\perp^{(\text{CM})}}{2\sigma_\perp^\gamma} \fr{1}{\ell_0+1}\right)^2 \ll 1,
\end{equation}
and the same behavior appears on the opposite side of the distribution ($m>0$).

Hence, the key parameter governing the OAM distribution in the evolved CM state is the ratio $\ell_0 b / \sigma_\perp^{(\text{CM})}$. In the regime $\ell_0 b / \sigma_\perp^{(\text{CM})} < 1$, the probability $P(\ell_0)$ dominates, and the mean OAM is primarily determined by the nearest-neighbor contribution $P(\ell_0-1)$:
\begin{equation}
    \expv{\ell} \approx \fr{\ell_0 P(\ell_0) + (\ell_0-1) P(\ell_0-1)}{P(\ell_0) + P(\ell_0-1)} \\
    = \ell_0 - \fr{(\ell_0 b / \sigma_\perp^{(\text{CM})})^2}{1 + (\ell_0 b / \sigma_\perp^{(\text{CM})})^2},
\end{equation}
with an OAM uncertainty $\Delta \ell \lesssim 1$.

In the opposite regime, $\ell_0 b  > \sigma_\perp^{(\text{CM})} \sim 1\; \text{nm} - 1 \;\mu$m, the mean value shifts toward the point where probabilities of neighboring OAM values become comparable, while the uncertainty remains $\Delta \ell \lesssim 1$:
\begin{equation}
    P(\ell_0 - m^*-1) \approx P(\ell_0 - m^*) 
     \Rightarrow \expv{\ell} \approx \fr{\ell_0}{1+  b / \sigma_\perp^{(\text{CM})}}.
\end{equation}}

\section{Comment on wave function shaping \label{App_shaping}}
Let us consider an important limiting case when the initial CM state is a plane wave propagating along $z$ axis with a definite momentum $P_{iz}$ and a \textit{paraxial} photon wave packet possesses a definite value of $k_z$, but is not necessarily twisted. It is not straightforward, however, to go to this limit from Eq.~\eqref{Transormed_Ev_state_CM}, because the decomposition of transverse delta function into angular delta functions in cylindrical coordinates has little sense once the transverse momentum is zero. Instead, we should return to the Eq.~\eqref{Transormed_Ev_state_CM} and assume the following:
\begin{equation}
    \delta^{(3)}(\bP_i+\bk- \bP_f) =  \delta^{(2)}(\bk_{\perp}- \bP_{f\perp})\delta(P_{iz}+k_{z}-P_{fz});
\end{equation}
\begin{gather}
    \psi_i^{\text{(CM)}}(\bP_{i}) = \delta^{(2)}(\bP_{\perp i})\delta\left(P_{iz}-\expv{P_{iz}}\right).\\
  \psi_i^{\gamma}(\bk) =  \Phi_i^{\gamma}(\bk_{\perp})\delta(k_z-\expv{k_z}). 
\end{gather}
The paraxial approximation is implied in the sense that $|\bk_\perp|\ll|k_z|$ and all the plane wave components of the photon packet have approximately the same polarization:
\begin{equation}
    \bm A_i(\bm r) \approx \bm \chi_{\lambda_i} e^{i\expv{k_z}z}\int \fr{d^2 \bk_\perp}{(2\pi)^2}e^{i\bk_\perp \bm r}\Phi_i^{\gamma}(\bk_{\perp}).
\end{equation}

Thus, the terms with $\bm e_{\lambda_i} \cdot \bP_i$ in electron integrals vanish because $\bm \chi_{\lambda_i}$ is orthogonal to $z$.
All the integration in Eq.~$(5)$ in this case is trivial due to delta functions, and we obtain
\begin{multline}
\la\bP_f|\psi^{\text{(CM, ev)}}(t_f)\ra 
    =-ie e^{i (m_i-m_f)\phi_f}e^{-i t_f \varepsilon^{\text{(CM)}}_f(P_f)} \fr{1}{(2\pi)^2}  \fr{1}{\sqrt{2\omega_k}} \delta\left(P_{fz}-\expv{k_z}-\expv{P_{iz}}\right)\\ 
    \times \delta \left(\varepsilon^{\text{(CM)}}_i(\bP_i)+\varepsilon^{e}_i+\omega_i-\varepsilon^{\text{(CM)}}_f(\bP_f)-\varepsilon^e_f\right)
     \Phi_i^{\gamma}(\bP_{f\perp})\tilde{\mathcal{J}}_{\lambda_i}(P_{f\perp}, P_{fz}),
\end{multline}
where 
\begin{equation}
   \tilde{\mathcal{J}}_{\lambda} = \sum_{m_i', m_f'}  d^{\ell_i}_{m_im_i'}(\theta_f) d^{\ell_f}_{m_fm_f'}(\theta_f)\left(-\fr{1}{m}\,{j}^{(1)}_{\lambda}-\fr{Z}{M}\,{j}^{(3)}_{\lambda}\right).
\end{equation}
Since due to the paraxiality $\theta_f\ll 1$ and $d^{\ell}_{mm'}(\theta_f)\approx \delta_{mm'}$, the obtained expression indicates that the CM wave function approximately acquires the transverse structure of the photon.

\section{Parametrization of $\dd^3 k$ integrals} \label{lucky}
The expression \eqref{Transormed_Ev_state_CM} for the evolved state of the CM after the photon absorption is convenient in terms of analytics: it allows one to discriminate the azimuthal phase - a hallmark of a twisted state. However, it turns out to have very limited applicability for numerical calculations because of the resonant nature of the absorption process. For the numerical integration, which is essential for calculation of probabilities, cross sections, etc., we propose an alternative parametrization of the integral~$(5)$ in the main text.
First, let us use the delta function to integrate over $\bP_i$, which yields $\bP_i = \bP_f - \bk$. The remaining delta function of energy can be rewritten as
\begin{multline}
       \delta\left(\varepsilon^{\text{(CM)}}_i(\bP_i)+\varepsilon^{e}_i+\omega-\varepsilon^{\text{(CM)}}_f(\bP_f)-\varepsilon^e_f\right)  \\
       = \frac{\delta\Big(\omega +(m+M) - |\bP_f|\cos \Theta - \sqrt{(m+M-|\bP_f|\cos \Theta)^2 + 2(m+M)(\varepsilon_f-\varepsilon_i)}\Big)}{\Big|\fr{\omega}{m+M} - \fr{|\bP_f|\cos\Theta}{m+M}+1\Big|},
\end{multline} 
where $\Theta$ is the angle between $\bP_f$ and $\bk$. The second root of the quadratic equation can be omitted because it yields a negative value of $\omega$. Thus, we get a rather simple two-dimensional integral over the photon angles $\theta_k$ and $\phi_k$. Most importantly, we have explicitly found a resonance condition for an arbitrary relative angle between the plane-wave components of photonic and atomic wave packets.

For scattering processes the kinematics becomes slightly more complex due to an additional photon in the final state. The parametrization, however, can be done is a similar way:

\begin{multline}
\delta\left(\varepsilon^{\text{(CM)}}_i(\bP_i)+\varepsilon^{e}_i+\omega_i-\varepsilon^{\text{(CM)}}_f(\bP_f)-\varepsilon^e_f - \omega_f \right)  \\
       = \delta\Bigg(\omega_i +(m+M) - |\bP_f|\cos \Theta_2 - |\bk_f|\cos \Theta_1 \\
       - \sqrt{(m+M-|\bP_f|\cos \Theta_2 - |\bk_f|\cos \Theta_1)^2 + 2(m+M)(\varepsilon_f-\varepsilon_i + \omega_f) - 2\omega_f |\bP_f|\cos \Theta_3 - \omega_f^2}\Bigg) \\
        / \Bigg|\fr{\omega_i}{m+M} - \fr{|\bP_f|\cos\Theta_2 + |\bk_f|\cos\Theta_1}{m+M}+1\Bigg|
\end{multline} 
with $\Theta_1$, $\Theta_2$ and $\Theta_3$ being the angles between vectors $\bk_i$ and $\bk_f$, $\bk_i$ and $\bP_f$, $\bk_f$ and $\bP_f$ respectively and momentum conservation $\bP_i + \bk_i = \bP_f + \bk_f$ is implied.

\section{Additional figures}
\label{appendix_plots}
\mbox{}

\begin{figure}[h]
	\centering
	\begin{subfigure}{0.485\linewidth}
		\includegraphics[width=\linewidth]{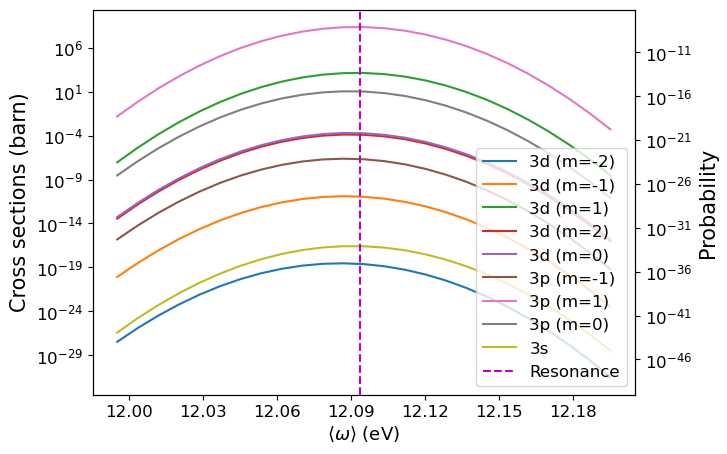}
        \caption{$\ell_\gamma=0$}
	\end{subfigure}
    \begin{subfigure}{0.485\linewidth}
    	\includegraphics[width=\linewidth]{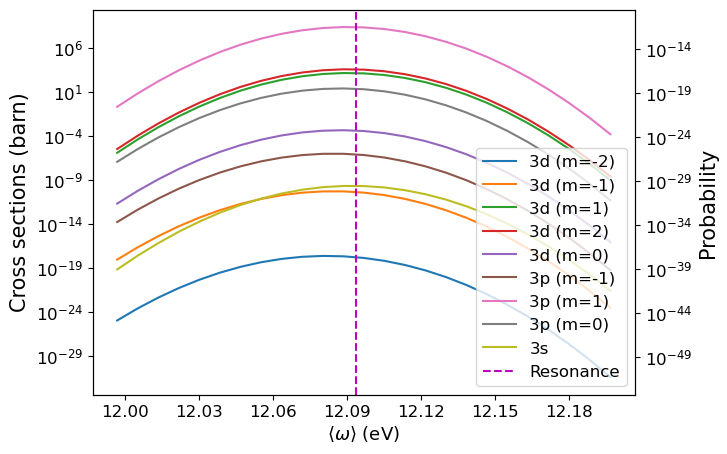}
        \caption{$\ell_\gamma=1$}
	\end{subfigure}

    \caption{$n=1\to n=3$ transitions in hydrogen: the probabilities and cross sections of absorption  versus the \textit{average} photon energy. The parameters of the wave packets are the same as in Fig. 2 in the main text. }
    \label{n1_to_n3}
\end{figure}

\begin{figure}[h]
	\centering
    \begin{subfigure}{0.475\linewidth}
    	\includegraphics[width=\linewidth]{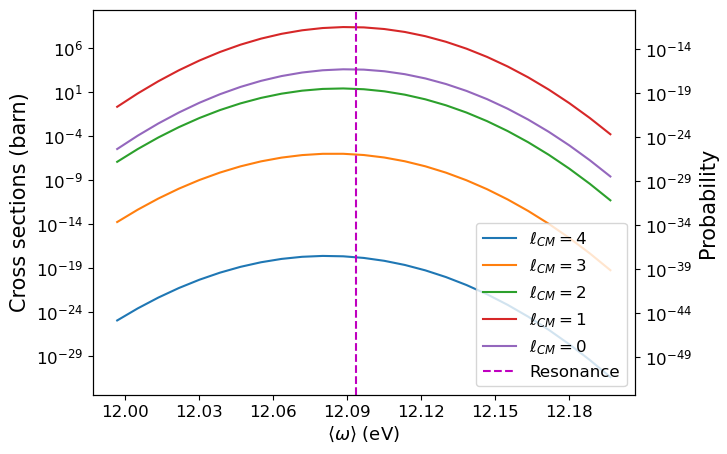}
        \caption{$\ell_\gamma=1$}
	\end{subfigure}
    \begin{subfigure}{0.475\linewidth}
    	\includegraphics[width=\linewidth]{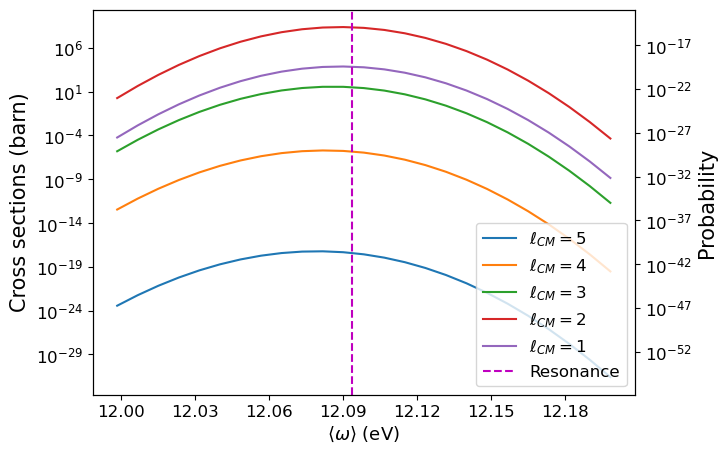}
        \caption{$\ell_\gamma=2$}
	\end{subfigure}

	\caption{$n=1\to n=3$ transitions: the cross sections of the OAM transfer from the twisted photon to the CM as a function of \textit{the mean} photon energy for the different OAM values. The dominant channel is always the one with $\ell^{\text{(CM)}} = \ell_\gamma$ due to the prevalence of dipole transitions. The parameters of the wave packets are the same as in Fig.~$2$ in the main text.}
    \label{CM_OAM}
\end{figure}

\begin{figure}[h]
    \centering
    \includegraphics[width=0.5\linewidth]{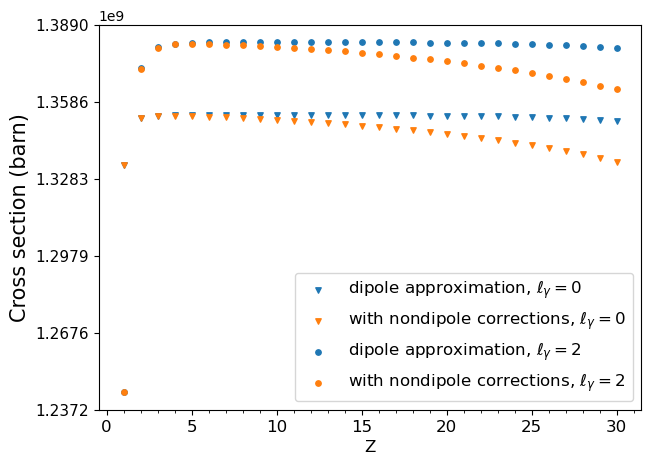}
    \caption{Cross section (15) for the dipole transition $1s \to 2p$ with $m_f = \lambda_i = 1$ at the resonance. Both the packets are gaussian, with $\sigma^\gamma_\perp = 1~\mu\text{m}$, $\sigma^\gamma_z = 10~\mu\text{m}$, and $\sigma^{\text{(CM)}}_\perp = \sigma^{\text{(CM)}}_z = 20~\text{nm}$. The nuclear charge $Z$ and mass number $A$ are varied jointly according to $A = 2Z$. The orange points show the calculated results without approximations, while blue points show the results within the dipole approximation for the transition currents.
}
    \label{Z_depend}
\end{figure}

\section{Nonparaxial effects}
\begin{figure}[h]
    \centering
    \includegraphics[width=0.5\linewidth]{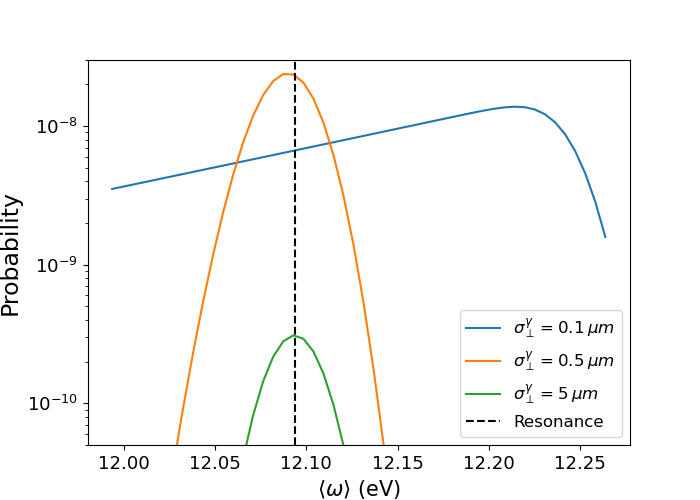}
    \caption{The asymmetry caused by the transverse coherence length of the photon: the probability (13) of the dipole transition $1s\to3p$ for different $\sigma^\gamma_\perp$. Both the packets are gaussian, $\sigma^\gamma_z = 10$ $\mu$m, $\sigma^{\text{(CM)}}_z = \sigma^{\text{(CM)}}_\perp=5$ nm. The black vertical line indicates the transition frequency of $12.1$ eV. }
    \label{nonparax_abs}
\end{figure}

\begin{figure}[h]
	\centering
    \includegraphics[width=0.5\linewidth]{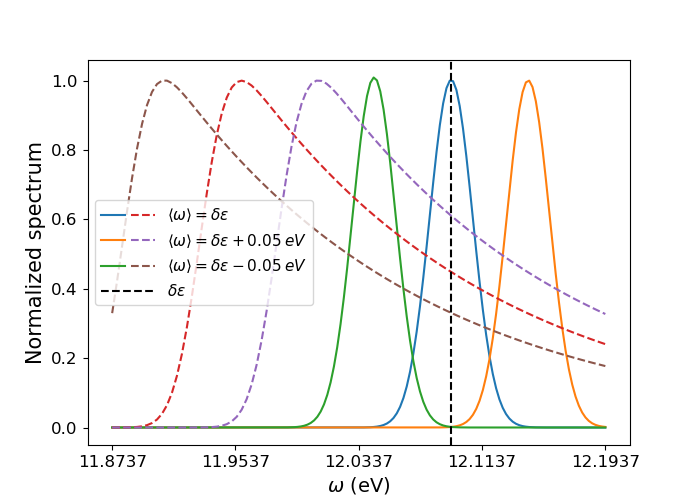}
	\caption{The energy spectra of the gaussian photonic packets with the different transverse coherence $\sigma^\gamma_\perp$, whereas the longitudinal   coherence length is fixed, $\sigma^\gamma_z = 10$ $\mu$m. Solid lines: paraxial packets with $\sigma_\perp^\gamma = 5\: \mu$m; dashed lines: nonparaxial packets with $\sigma_\perp^\gamma = 0.1\: \mu$m. The spectra are calculated by averaging the energy over all the angles with the momentum-space probability density of the photon. The spectrum is normalized to a dimensionless form with a unit maximum. The black vertical line indicates the transition frequency $\delta \varepsilon = \varepsilon^e_f - \varepsilon^e_i$.}
\label{nonpar_spectrum}
\end{figure}

 Results in the main text are presented only for paraxial photons, that is, those with a large $\sigma^\gamma_\perp$, such that $1 / \sigma^\gamma_\perp \ll \expv{\omega}$. In the nonparaxial regime, the hierarchy of transition probabilities stays qualitatively the same. The \textit{asymmetry} of the absorption line, however, increases with the decrease of the photon transverse coherence: for the average frequencies above the resonance, the cross section still drops rapidly, while below the resonance the decay becomes slower. This is explained by the influence of nonparaxiality on the photon spectrum, which is shown in Fig. \ref{nonparax_abs}. For tightly focused photons the dispersion of transverse momentum can become comparable to the photon mean energy. Thus, the spectrum itself becomes strongly {\it asymmetric} due to the increased contribution of frequencies above the average one. As a result, a small shift of the average frequency does not substantially change the amplitude of the plane-wave components that are at a resonance with the electronic transition, which is shown in Fig. \ref{nonpar_spectrum}. In the opposite paraxial case with $\sigma^\gamma_\perp \gg 1 / \expv{\omega}$, the spectrum is narrow and almost symmetric, and even a small shift of the mean energy from the spectral maximum leads to a significant decrease of the amplitude, as demonstrated also in Fig. \ref{nonpar_spectrum}.

\section{The role of longitudinal wave function of the photon}

In addition to the transverse coherence length and the OAM, one can also control the longitudinal structure of the photon packet. The HG wave functions from Eq.~\eqref{HG_LG_phot}, similar to those of a one-dimensional harmonic oscillator, correspond to optical pulses with {\it a finite number of peaks in the longitudinal direction} (both in real space and in the momentum one), controlled by a quantum number $k_\gamma = 0, 1, 2,...$. Shaping the longitudinal wave functions affects the energy spectrum of the photonic packet by adding $k_\gamma$ peaks to the energy distribution. 

\begin{figure}[h]
	\centering
    \includegraphics[width=0.5\linewidth]{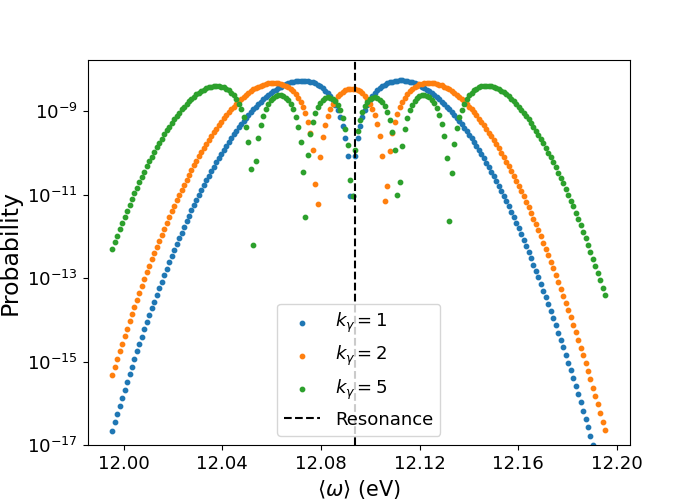}

	\caption{The probability (13) of the dipole transition $1s\to 2p, m_f=1$ for $\lambda_i=1$ as a function of the mean photon energy for the different longitudinal HG function of the photon with $k_\gamma = 0, 1, 2,...$  being an integer quantum number from Eq.~\eqref{HG_pos}. The transverse photonic wave function is gaussian with $\sigma^\gamma_\perp = 1$ $\mu$m, the longitudinal coherence length is $\sigma^\gamma_z = 10$ $\mu$m, which corresponds to a femtosecond laser pulse. The mean energy of the photon exactly matches the transition energy $\varepsilon^e_f-\varepsilon^e_i$.}
    \label{HG_split}
\end{figure}

Similarly to the line asymmetry caused by the spectrum skew, here the absorption lines inherit the $k_\gamma+1$-peak shape, as demonstrated in Fig. \ref{HG_split}. The corresponding spectra of the longitudinally structured photonic packets with the gaussian transverse profile are presented in Fig. \ref{HG_spec}. By analogy with the effect of transverse shaping of atomic packet, discussed in Sec. IX of Supplementary Materials, we expect that the longitudinal structure of the photonic packet {\it can also be ``imprinted'' on the CM wavefunction}, giving rise to its modulation along the collision axis.

\begin{figure}[h]
    \centering
    \includegraphics[width=0.5\linewidth]{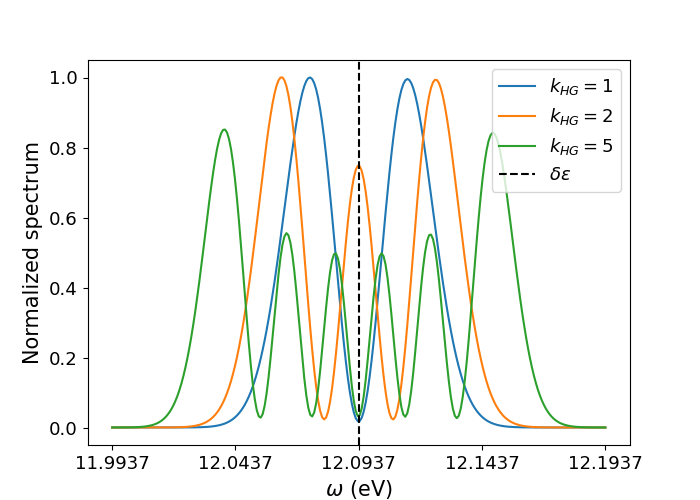}
    \caption{The energy spectra of the photonic HG packets from Eq.~\eqref{HG_pos} with different longitudinal quantum number $k_\gamma$ controlling the number of peaks. Parameters are the same as in Fig. \ref{HG_split}.}
    \label{HG_spec}
\end{figure}

\bibliographystyle{unsrt}
\bibliography{ref}